\documentclass[12pt]{article}
\pdfoutput=1
\usepackage{amsfonts}
\usepackage{amsmath}
\usepackage{url}
\usepackage{hyperref}
\usepackage{bbold}
\usepackage{slashed}
\usepackage{tikz}
\usepackage{graphicx}
\usepackage{epstopdf}
\usepackage{subfigure}
\usepackage{pgfplots}
\usepackage{amssymb}
\usepackage{color}
\usepackage{mathrsfs}
\usepackage{fancybox}
\usepackage{lipsum}
\usepackage{eurosym}
\usepackage{tcolorbox}
\usepackage{physics}
\usepackage{tensor}

	\newcommand{\LL}{\scriptscriptstyle \text{L}}

\newcommand{\mo}{\mathcal{O}}
\newcommand{\mm}{\mathcal{M}}

\newcommand{\airy}{\text{Ai}}

\newcommand{\disk}{\text{Disk}}

\newcommand{\Poincare}{Poincar\'e }
\newcommand{\sinc}{\text{sinc}}
\newcommand{\average}[1]{\left\langle #1 \right\rangle}

\begin{document}
\begin{titlepage}

\setcounter{page}{1} \baselineskip=15.5pt \thispagestyle{empty}

\vfil

${}$
\vspace{1cm}

\begin{center}
\def\thefootnote{\fnsymbol{footnote}}
\begin{changemargin}{0.05cm}{0.05cm} 
\begin{center}
{\Large \bf Eigenbranes in Jackiw-Teitelboim gravity}
\end{center} 
\end{changemargin}

~\\[1cm]
{Andreas Blommaert\footnote{\href{mailto:andreas.blommaert@gmail.com}{\protect\path{andreas.blommaert@gmail.com}}}, Thomas G. Mertens\footnote{\href{mailto:thomas.mertens@ugent.be}{\protect\path{thomas.mertens@ugent.be}}} and Henri Verschelde\footnote{\href{mailto:henri.verschelde@ugent.be}{\protect\path{henri.verschelde@ugent.be}}}
}
\\[0.3cm]

{\normalsize { \sl Department of Physics and Astronomy
\\[1.0mm]
Ghent University, Krijgslaan, 281-S9, 9000 Gent, Belgium}}\\[3mm]

\end{center}


 \vspace{0.2cm}
\begin{changemargin}{01cm}{1cm} 
{\small  \noindent 
\begin{center} 
\textbf{Abstract}
\end{center} 
It was proven recently that JT gravity can be defined as an ensemble of $L\times L$ Hermitian matrices. We point out that the eigenvalues of the matrix correspond in JT gravity to FZZT-type boundaries on which spacetimes can end. We then investigate an ensemble of matrices with $1\ll N\ll L$ eigenvalues held fixed. This corresponds to a version of JT gravity which includes $N$ FZZT type boundaries in the path integral contour and which is found to emulate a discrete quantum chaotic system. In particular this version of JT gravity can capture the behavior of finite-volume holographic correlators at late times, including erratic oscillations.
}

\end{changemargin}
 \vspace{0.3cm}
\vfil
\begin{flushleft}
\today
\end{flushleft}

\end{titlepage}

\newpage
\tableofcontents
\vspace{0.5cm}
\noindent\makebox[\linewidth]{\rule{\textwidth}{0.4pt}}
\vspace{1cm}

\addtolength{\abovedisplayskip}{.5mm}
\addtolength{\belowdisplayskip}{.5mm}

\def\plus{\raisebox{.5pt}{\tiny$+$\smpc}}

\addtolength{\parskip}{.6mm}
\def\spc{\hspace{1pt}}

\def\nspc{{\hspace{-2pt}}}
\def\ff{\rm\smpc f\smpc} 
\def\fff{\mbox{Y}}
\def\ww{{\rm w}}
\def\smpc{{\hspace{.5pt}}}

\def\zz{{\spc \rm z}}
\def\xx{{\rm x\smpc}}
\def\xxi{\mbox{\footnotesize \spc $\xi$}}
\def\jj{{\rm j}}
\addtolength{\baselineskip}{-.1mm}

\renewcommand{\Large}{\large}

\setcounter{tocdepth}{2}
\addtolength{\baselineskip}{0mm}
\addtolength{\parskip}{.4mm}
\addtolength{\abovedisplayskip}{1mm}
\addtolength{\belowdisplayskip}{1mm}

\setcounter{footnote}{0}

\section{Introduction}

In finite-volume holography, there is a deep tension between discreteness of the boundary theory, and quasi-normal decay in the holographic bulk. This tension is one version of the information paradox, due to Maldacena \cite{0106112}. A bulk quantum gravity explanation for this behavior remains to some degree an open question, though important steps have been taken in \cite{bhrm,semiclassicalramp,sss2}. Moreover in general it is not yet completely clear how semiclassical gravitational arguments are to be augmented in a way that resolves the information paradox. For recent progress see though \cite{1905.08255,1905.08762,1908.10996,1910.00972}.
\\
Two-dimensional Jackiw-Teitelboim (JT) gravity \cite{jackiw,teitelboim} has attracted a lot of attention in recent years. This is largely due to its exact solubility and its relevance as the low-energy universal sector of the SYK model \cite{Maldacena:2016hyu,Almheiri:2014cka,1605.06098,1606.01857,1606.03438}. JT gravity comes in different versions, as we will highlight in section \ref{sect:discrete}. Most activity thus far has been in a version of JT gravity dual to Schwarzian quantum mechanics, which comes with a topologically trivial bulk. The Schwarzian has a continuous spectrum, so this version of JT gravity has fairly little to do with holographic discreteness.
\\
More recently there has been interest in a version of JT gravity that includes a sum over higher genus topologies in the bulk \cite{sss2}. This model can be defined non-perturbatively as a double-scaled matrix integral. It is therefore an ensemble average over discrete systems. Due to the averaging, the spectrum remains continuous. Nonetheless the averaging does not eradicate all traces of discreteness, in particular late-time holographic correlators do not generically decay to zero in this version of JT gravity \cite{sss2,paper5,phil,Cotler:2019egt}.
\\
Motivated by these developments, we want to point out that there exists a further alternative version of JT gravity which resembles the behavior of a single discrete holographic system. In particular it captures the behavior of the spectral form factor of a discrete system for all times, including the late-time erratic oscillations. The new feature is to include a of set fixed energy boundaries in the gravitational path integral on which Riemann surfaces can end, to be distinguished from the asymptotic boundaries. Each energy label corresponds to the energy of a state in the discrete system we aim to emulate. These boundaries, which we will refer to as eigenbranes, are related to unmarked FZZT boundaries and correspond to fixed eigenvalues in the matrix ensemble of \cite{sss2}.\footnote{FZZT boundaries in Liouville theory were discussed in \cite{0001012,teschner}. Our eigenbranes are related to FZZT boundaries in the following way. They are the products of exponentials of unmarked FZZT boundaries as discussed in \cite{sss2} and in this context in appendix \ref{app:calculations}: in that sense eigenbranes are really branes (exponentials of boundaries). This choice of Liouville language comes from the link between JT gravity as the $p\to \infty$ limit of the $(2,p)$ minimal string, as discovered in \cite{sss2}.}
\\~\\
This work is structured as follows.\\
In \textbf{section \ref{sect:discrete}} we introduce a simple probe of discreteness. Using the freedom to choose the contour of the path integral over metrics in quantum gravity, we discuss three possible definitions of JT gravity. One has only disk topologies, one has all topologies and is completed as an ensemble of Hermitian random matrices. The final one includes eigenbranes and is dual to an ensemble of Hermitian random matrices with a large number of eigenvalues fixed.\\
In \textbf{section \ref{sect:preliminary}} we review and discuss calculational techniques for spectral densities in matrix integrals, or multi-boundary correlators in gravity.\\
In \textbf{section \ref{sect:fixing}} we consider an ensemble of random matrices with certain eigenvalues kept fixed, and derive its perturbative interpretation as a gravity path integral with surfaces ending also on a number of fixed-energy boundaries. We prove the extent to which the resulting version of JT gravity resembles a discrete system. In particular, we show that the spectrum essentially reduces to a series of delta functions, and we show how different asymptotic boundaries essentially disconnect.\\
In \textbf{section \ref{sect:concl}} we briefly discuss a possible gravitational interpretation of the delta functions as due to boundary mergers \cite{1911.01659} and touch on the generalization to JT supergravity \cite{stanfordwittenJT}.\\
The \textbf{appendices} contain some of the technical material.

\section{Diagnosing discreteness}\label{sect:discrete}
Consider a particular discrete maximally chaotic quantum mechanical system with an $L$-dimensional discrete Hilbert space:
\begin{equation}
    \rho(E)=\sum_{i=1}^L \delta(E-\lambda_i).\label{spikes}
\end{equation}
We choose the model in such a way that its coarse-grained spectrum matches the JT spectrum on the disk up to some large energy cutoff $\Lambda\gg 1$ \cite{Maldacena:2016hyu, 1606.01857,bhrm,Stanford:2017thb}:\footnote{We have set the dimensionful quantity $C=1/2$ which has units of length and comes from the symmetry breaking mechanism in NAdS$_2$/NCFT$_1$. All energy scales are refered to this scale.}
\begin{equation}
    \rho_{\text{coarse}}(E)=\frac{e^{S_0}}{4\pi^2}\sinh 2\pi \sqrt{E}=\rho_0(E),\qquad E<\Lambda.\label{lambda}
\end{equation}
The system is taken to be quantum chaotic, because we aim for it to be dual to quantum black holes \cite{bhrm,1503.01409,1306.0622,1412.6087}. Specifying further to a system without time-reversal invariance, this implies its local level statistics should be those of a random matrix taken from an appropriately rescaled Gaussian unitary ensemble (GUE) \cite{mehta}.
\\
At low energies $E < \Lambda$ or late times, we might expect that this discrete quantum mechanical system has an effective pure JT gravity bulk description.\footnote{The late-time behavior of chaotic systems was recently studied in \cite{Cotler:2019egt}, where it was found that late-time correlation functions factorize into a purely spectral quantity governing the time-dependence, and an operator-dependent prefactor. This spectral quantity is probe-independent and expected to have a pure gravity bulk description.} In the remainder of this work we collect evidence in favor of this.
\\~\\
We will be interested in thermal correlation functions, for example the two-point function:\footnote{It is conventional to rename $\beta \to 2 \beta$.}
\begin{align}
		\nonumber \Tr[ e^{-\beta H}\mathcal{O}(t)\,e^{-\beta H}\mathcal{O}(0)] &=\\ \int_{-\infty}^{+\infty} d E_1\, \rho(E_1)\,& e^{-(\beta+it) E_1 }\int_{-\infty}^{+\infty} d E_2\, \rho(E_2)\, e^{-(\beta-it)E_2 }\,\abs{\bra{E_1}\mo \ket{E_2}}^2.\label{twopoint}
\end{align}
At early times, the Fourier transform is insensitive to the fine structure of the spectrum, and coarse-grains. As a consequence, the thermal two-point function will be well approximated by a JT disk calculation. The late-time Fourier transform on the other hand is highly sensitive to the fine structure. Therefore late-time correlators are in general suitable probes of discreteness \cite{0106112}.
\\
In particular we will be interested in the simplest such probe, a local version of the two-point function where we integrate both $E_1$ and $E_2$ over only a narrow energy interval $\text{bin}(E)$. We take $1/\rho_0(E)\ll\abs{\text{bin}(E)}\ll 1$, this is small enough such that $\rho_0(E)$ is approximately constant, and large enough such that it contains a large number of eigenvalues. Because the system \eqref{spikes} is quantum chaotic, we known from the eigenvalue thermalization hypothesis (ETH) that $\abs{\bra{E_1}\mo \ket{E_2}}^2$ are also slowly varying in this bin \cite{eth1,eth2,phil,bhrm}:
\begin{equation}
    \abs{\bra{E}\mo \ket{E}}^2\int_{\text{bin}(E)}dE_1\, \rho(E_1)\, e^{-(\beta+it) E_1 }\int_{\text{bin}(E)} d E_2\, \rho(E_2)\, e^{-(\beta-it)E_2 }.
\end{equation}
We are interested only in the $t$-dependence of this object, so we can remove the constant matrix element and $\beta$ dependence (we are not interested in $\beta\gg 1$ hence the $\beta$ dependence comes out as a prefactor). We are left with a local version of the spectral form factor \cite{semiclassicalramp}:
\begin{equation}
    S_E(t)=\int_{\text{bin}(E)}dE_1\, \rho(E_1)\, e^{-it E_1 }\int_{\text{bin}(E)} d E_2\, \rho(E_2)\, e^{it E_2 }.\label{spectralmicro}
\end{equation}
Labeling the eigenvalues of \eqref{spikes} within $\text{bin}(E)$ as $\lambda_1\dots \lambda_N$, with $1\ll N\ll L$, we get:
\begin{equation}
  S_E(t)=\sum_{i,j=1}^N \cos t(\lambda_i-\lambda_j) = N + \sum_{i\neq j}^N \cos t(\lambda_i-\lambda_j).\label{tofind}
\end{equation}
A log-log plot for a representative sample gives:\footnote{We took $128$ consecutive eigenvalues of one large matrix drawn from a GUE far from the edge.}
\begin{equation}
    S_E(t)=\quad\raisebox{-27mm}{\includegraphics[width=0.7\textwidth]{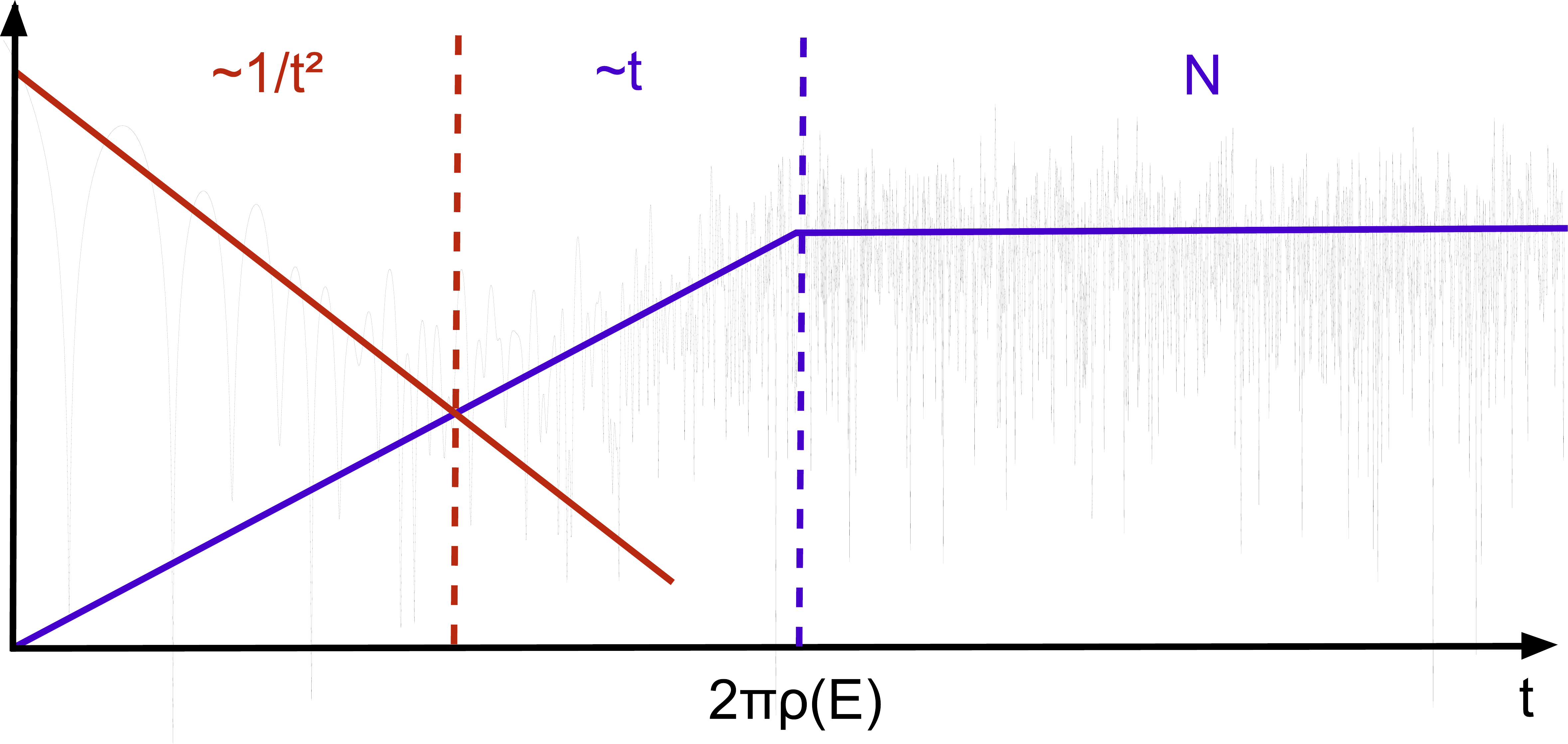}}\label{plotsff}
\end{equation}
As compared to the usual spectral form factor, there are additional low-frequency oscillations, but the general shape is the same \cite{bhrm}. In particular, it has the same ramp-and plateau structure including erratic oscillations, with plateau height $N$. We will reproduce these erratic oscillations via a bulk JT gravity calculation in section \ref{sect:erratic}.

\subsection{Models of JT gravity}
JT gravity is a model of 2d dilaton gravity with action \cite{jackiw,teitelboim}:
\begin{equation}
    S_\text{JT}[g,\Phi]=-\frac{\Phi_0}{4G}\chi(\mathcal{M}) -\frac{1}{2}\int_\mm d^2 x \sqrt{g} \Phi (R+2)-\int_{\partial \mm}d t \sqrt{h} \Phi (K-1).
\end{equation}
The Euler character $\chi(\mathcal{M})$ comes from the Einstein-Hilbert term in 2d.\footnote{For a 2d manifold of genus $g$ with $b$ boundaries we have $\chi=2-2g-b$.} The quantity $S_0 = \Phi_0/4G$ corresponds to the extremal entropy, and is a free parameter from the gravity point of view. 
\\
Integrating out $\Phi$ localizes the metrics $g$ on hyperbolic Riemann surfaces, or patches of the \Poincare disk, with asymptotically NAdS$_2$ boundary conditions. This boils down in JT gravity to fixing the total length of the asymptotic boundary to $\beta/\epsilon$, and the boundary value of the dilaton to $1/2\epsilon$ \cite{1605.06098,1606.01857,1606.03438}. Defining the integration space of $g$ boils down to specifying which surfaces to count. For the JT gravity partition function, which corresponds to the insertion of a single holographic boundary in the path integral, we have schematically:
\begin{equation}
    Z_\text{JT}(\beta)=\int d E\, e^{-\beta E}\, \rho(E)=\quad \beta\, \raisebox{-9mm}{\includegraphics[width=15mm]{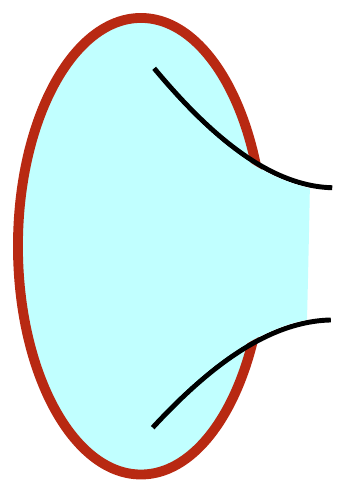}}???\label{what}
\end{equation}
The spectral form factor is related in JT gravity to a correlation function $Z_\text{JT}(\beta_1,\beta_2)$ with two asymptotic boundaries of respective lengths $\beta_1/\epsilon$ and $\beta_2/\epsilon$:
\begin{equation}
    Z_\text{JT}(\beta_1,\beta_2)=\int dE_1\,e^{-\beta_1E_1}\int dE_2\,e^{-\beta_2E_2}\,\rho(E_1,E_2)=\quad \beta_1\, \raisebox{-8.5mm}{\includegraphics[width=14mm]{what.pdf}}???\raisebox{-8.5mm}{\includegraphics[width=14mm]{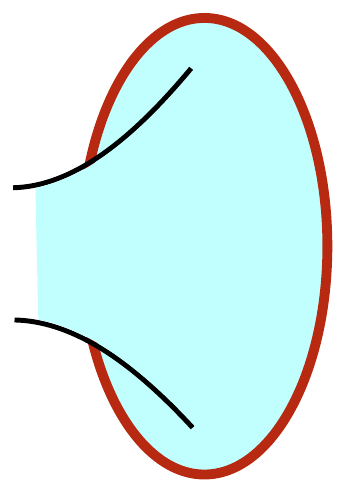}}\,\beta_2\label{zb1b2}
\end{equation}
From this, one calculates the local spectral form factor \eqref{spectralmicro} in gravity as:
\begin{equation}
    S_E(t)=\int_\text{bin(E)} d E_1\, e^{it E_1}\int_\text{bin(E)} d E_2\, e^{-it E_2}\rho(E_1,E_2).\label{sff}
\end{equation}
In the remainder of this section we highlight how different definitions of $???$ can lead to structurally very different theories, using the spectral form factor \eqref{sff} as probe. The different models we will discuss can be specified by the integration space in the path integral over metrics:
\begin{align}
\int_{\text{disks}}\left[\mathcal{D}g\right] (\hdots)\quad, \qquad \int_{\text{all $\chi$}}\left[\mathcal{D}g\right] (\hdots)\quad , \qquad \int_{\lambda_1\dots\lambda_n\, \text{all $\chi$}}\left[\mathcal{D}g\right] (\hdots)\quad.\label{dg}
\end{align} 
In particular, we want to point out that the last definition which takes the energies $\lambda_1\dots \lambda_n$ as input, resembles the discrete system with spectrum \eqref{spikes}.

\subsubsection{Version 1: Disks}
The simplest definition is to restrict $???$ to Riemann surfaces which are topologically disks. The spectrum and correlation functions of JT gravity on the disk have been extensively studied in recent years, resulting in several complementary perspectives \cite{altland,altland2,Mertens:2017mtv,Mertens:2018fds,Lam:2018pvp,paper3,kitaevsuh,zhenbin,Mertens:2019tcm,1905.02726}. Its spectrum is continuous \cite{Maldacena:2016hyu, 1606.01857,bhrm,Stanford:2017thb}:
\begin{equation}
    \rho_0(E)=\frac{e^{S_0}}{4\pi^2}\sinh 2\pi \sqrt{E}.\label{diskspec}
\end{equation}
To probe the spectrum, we focus on the spectral form factor \eqref{spectralmicro}. For disk topologies, the gravitational spectral form factor factorizes $\rho(E_1,E_2)=\rho_0(E_1)\rho_0(E_2)$, and we get:
\begin{align}
\label{powerlaw}
S_E(t) = \rho_0(E)^2\int_\text{bin(E)} d E_1\, e^{it E_1}\int_\text{bin(E)} d E_2\, e^{-it E_2} = \frac{4\rho_0(E)^2}{t^2} \sin^2 \frac{\text{bin}(E)}{2} t .
\end{align}
This dependence correctly captures the part of the curve \eqref{plotsff} before the dotted red line, including the relatively slow oscillations. At later times, the Fourier transform is able to distinguish the coarse-grained disk spectrum \eqref{diskspec} from the discrete spectrum \eqref{spikes}. 

\subsubsection{Version 2: Genus expansion and random matrices}\label{sect:matrix}
A second possible definition of JT gravity is to allow for Riemann surfaces of arbitrary genus to end on the asymptotic boundaries. This version was introduced and discussed in \cite{sss2}.\footnote{Aspects of JT gravity on higher genus Riemann surfaces were also discussed in \cite{paper4,paper5,1904.01911,1905.03780,stanfordwittenJT,phil}. Let us note that an alternative mathematically consistent definition corresponds to integrating over Teichm\"uller space instead of the moduli space of Riemann surfaces \cite{paper4}. This is more natural from a first-order BF point of view, but it gives divergences at higher genus and hence is not a workable version of JT gravity as sum over topologies.}  The JT gravity partition function now has a genus expansion:
\begin{equation}
    Z_\text{JT}(\beta)=\int d E\, e^{-\beta E}\, \rho(E)=\quad \beta\, \raisebox{-9mm}{\includegraphics[width=12mm]{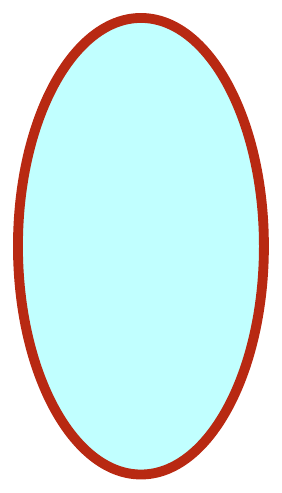}}\quad +\quad \beta \,\raisebox{-9mm}{\includegraphics[width=16.5mm]{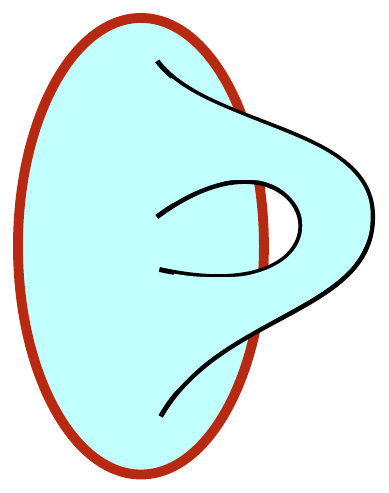}}\quad+\dots\label{genusexp}
\end{equation}
The same is true for the spectral density, and all other observables:
\begin{equation}
    \average{\rho(E)} = \rho_0(E) + \sum_{g=1}^\infty e^{-2 g S_0}\rho_g(E).
\end{equation}
We will refer to \eqref{genusexp} as the ``perturbative" definition of JT gravity.\footnote{It is perturbative in the string coupling $e^{-S_0}$ but non-perturbative in the Newton constant $G\sim 1/S_0$. Indeed, JT gravity is a string theory, much like string theory can generically be viewed as 2d quantum gravity on the worldsheet.} It is very feasible to calculate each term in this series, as we briefly review in section \ref{sect:31}. The resulting perturbative series turns out to be asymptotic, and hence requires a non-perturbative definition. We can define JT gravity non-perturbatively as a double-scaled matrix integral with genus zero spectral density \eqref{diskspec} \cite{sss2}.\footnote{see also \cite{0407261,1804.03275}.} In the weakly coupled regime $e^{S_0}\gg1$, it turns out one can neglect all perturbative contributions, and that the leading correction is nonperturbative in the string coupling:\footnote{This is only true when $E\gg e^{-2S_0/3}$. We will operate under this assumption throughout the main text. Otherwise, we have to resort to an exact analysis of the Airy model, as we do in appendix \ref{app:airy}.}
\begin{align}
    \average{\rho(E)} = \rho_0(E)+\rho_\text{nonp}(E), \qquad 
    \rho_\text{nonp}(E)= -\frac{1}{4\pi E}\cos(2\pi\int_0^E d M\rho_0(M)).
		\label{wiggle}
\end{align}
Such oscillatory non-perturbative contributions will in general not have a geometrical interpretation as counting Riemann surfaces.\footnote{The nonperturbative contribution in the forbidden region $E<0$ does seem to count Riemann surfaces that end on the appropriate boundaries, stretching between asymptotic boundaries and ZZ branes in the bulk \cite{sss2}. These are to be distinguished from the FZZT branes discussed in the remainder of this work.}
\\~\\
The partition function of an ensemble of $L\times L$ Hermitian matrices with bare potential $V(M)$ is defined as:
\begin{equation}
    \mathcal{Z}_L=\int D M e^{-L\Tr V(M)}.\label{partition}
\end{equation}
A more convenient way to write this is in terms of the eigenvalues $\lambda_i$ of the matrices:\footnote{Any contour $\mathcal{C}$ represents a definition of a matrix model. In case of JT gravity, for stability reasons the contour cannot be chosen to extend along the real energy axis all the way up to $-\infty$ \cite{sss2}. The part of the contour that deviates from the negative real axis is not important for the content of this work though, so we will drop the subscript in what follows. See appendix \ref{app:airy} for fixed eigenvalues in the forbidden region.}
\begin{equation}
    \mathcal{Z}_L=\int_\mathcal{C} \prod_{i=1}^L\left( d \lambda_i\, e^{-L V(\lambda_i)}\right) \Delta(\lambda_1,\dots,\lambda_L) , \qquad \Delta(\lambda_1,\dots,\lambda_L) = \prod_{i<j}^L(\lambda_i-\lambda_j)^2.\label{ensemble}
\end{equation}
Here $\Delta(\lambda_1,\dots,\lambda_L)$ is the Vandermonde determinant, accounting for eigenvalue repulsion. An intuitive way to think about about such matrix integrals is as the steady state of the Brownian motion of $L$ charged particles in an external potential $V(x)$, a so-called Dyson gas \cite{dyson1,dyson2,dyson3,dyson4,dyson5,dysonbrown}.
The Vandermonde determinant then represents the electrostatic repulsion. \\
Typical observables in the matrix model are products of the spectral density $\rho(E)$ or the macroscopic loop operator $Z(\beta)$:
\begin{equation}
    \rho(E)=\sum_{i=1}^L\delta(E-\lambda_i), \qquad  Z(\beta)=\sum_{i=1}^L e^{-\beta \lambda_i}.
\end{equation}
Correlators are calculated as ensemble averages, for example:\footnote{They are normalized such that $\average{\mathbf{1}} = 1$.}
\begin{equation}
    \average{Z(\beta)}=\frac{1}{\mathcal{Z}_L}\int d\lambda_1 \dots e^{-LV(\lambda_1\dots)}\Delta(\lambda_1\dots)\sum_{i=1}^L e^{-\beta \lambda_i}\label{zb}.
\end{equation}
The two-level spectral density is defined as $\rho(E_1,E_2) = \rho(E_1)\rho(E_2)$. Ensemble averaging leads to correlation as connected contributions, for example for the two-loop operator defined as $Z(\beta_1,\beta_2) = Z(\beta_1)Z(\beta_2)$:
\begin{equation}
    \average{Z(\beta_1,\beta_2)}=\average{Z(\beta_1)}\average{Z(\beta_2)}+\average{Z(\beta_1,\beta_2)}_\text{conn}.
\end{equation}
Comparing to the perturbative JT gravity definition of $Z(\beta_1,\beta_2)$ in \eqref{zb1b2}, which counts all Riemann surfaces that end on the two asymptotic boundaries, one sees that connected correlators correspond to connected geometries:
\begin{equation}
    \nonumber\average{Z(\beta_1,\beta_2)}_\text{conn} 
		= \quad  \beta_1\,\raisebox{-10mm}{\includegraphics[width=38mm]{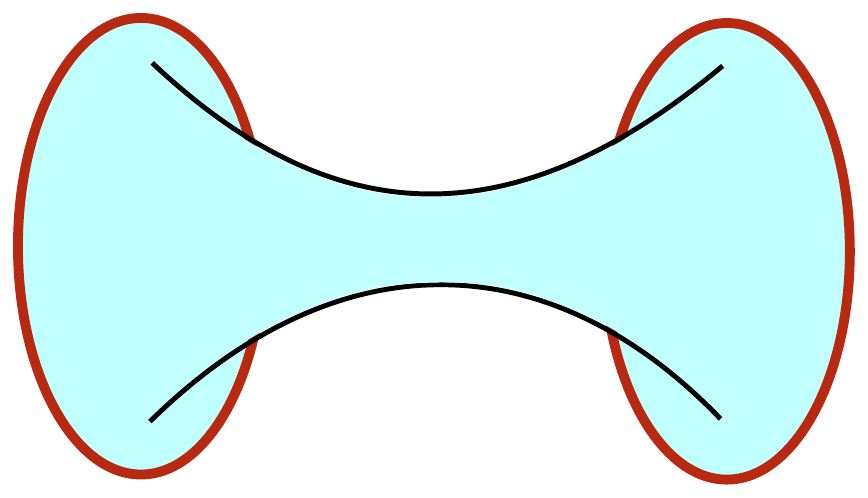}}\,\beta_2\quad+\quad \beta_1\,\raisebox{-10mm}{\includegraphics[width=38mm]{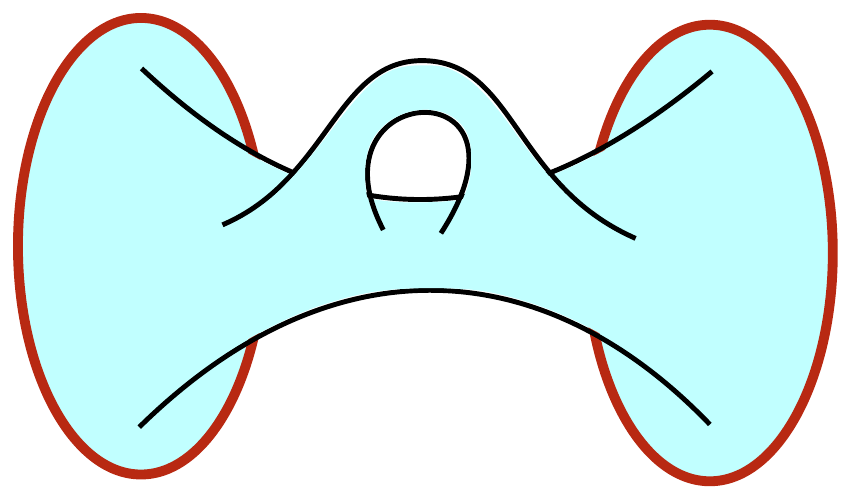}}\,\beta_2\quad +\dots\label{conngeom}
\end{equation}
Again, it is not hard to calculate both perturbative and nonperturbative contributions to $\average{\rho(E_1)\rho(E_2)})$, see section \ref{sect:preliminary}, appendix \ref{app:calculations} and \cite{sss2}. The only significant perturbative contributions are due to the disconnected disks, and annulus connecting the two boundaries. There are also significant nonperturbative contributions:\footnote{In addition to these contributions we find a generalization of the wiggles \eqref{wiggle} in appendix \ref{app:calculations}. Such wiggles are genuinely small corrections though, unlike \eqref{nptwo} and are not relevant for our discussion.}
\begin{equation}
    \average{\rho(E_1)\rho(E_2)} = \delta(E_1-E_2)\rho_0(E_1) + \rho_0(E_1)\rho_0(E_2) - \frac{1}{2\pi^2}\frac{1}{(E_1-E_2)^2}+\rho_\text{nonp}(E_1,E_2),\label{twolevel0}
\end{equation}
with
\begin{equation}
    \rho_\text{nonp}(E_1,E_2) = \frac{1}{2\pi^2}\frac{1}{(E_1-E_2)^2}\cos(2\pi \int_{E_1}^{E_2} d M\, \rho_0(M)).\label{nptwo}
\end{equation}
The second contribution here is of the same oscillatory type as \eqref{wiggle}. Unlike that contribution though, it isn't particularly small due to the multiplicative pole, and cannot be neglected in our analysis. 
\\~\\
The spectral form factor is calculated as in \eqref{sff}. The $ \rho_0(E_1)\rho_0(E_2) $-contribution in \eqref{twolevel0} gives the power-law decay \eqref{powerlaw}. The Dirac-delta yields the constant plateau contribution $N$, and the other contributions add up to the sine-kernel \eqref{cluster} which gives a variant of the ramp at late times:\footnote{In more detail, the contribution of the sine kernel is:
\begin{equation}
    S_E(t)\supset -N \int d\tau \left(\frac{1}{\pi}\frac{N}{2\rho(E)}\sinc^2 \frac{N}{2\rho(E)} \tau \right)\text{Ramp}\left(\frac{t-\tau}{2\pi\rho(E)}\right), \qquad \sinc(x) \equiv \frac{\sin(x)}{x}.\label{averagedramp}
\end{equation}
This is a low-frequency filtered version of the usual ramp. Qualitatively, at $t\ll 2\pi \rho(E)$ there will be significant smoothening of the onset of the usual ramp. In the regime of interest where $N \gg 1$, the kernel acts as a Dirac-function and one obtains the linear ramp with plateau time $t_{\text{plateau}} = 2\pi \rho(E)$ .}
\begin{equation}
    S_E(t)\supset N-N \text{Ramp}\left(\frac{t}{2\pi\rho(E)}\right), \qquad \text{Ramp}(x) = (1-\abs{x})\theta(1-\abs{x}).
\end{equation}
This resulting function follows \eqref{plotsff} before the dotted red line, and the blue curve at late times.

\subsubsection{Version 3: Eigenbranes}\label{sect:213}
The version of JT gravity discussed above is a double-scaled matrix integral. This means we take $L\to\infty$ and simultaneously zoom in on a region $E<\Lambda$ near the edge of the spectrum, keeping the total average number of eigenvalues $1\ll N\ll L$ in this region fixed. We can visualize this as:
\begin{equation}
    \average{\rho(E)}=\quad\raisebox{-9mm}{\includegraphics[width=0.3\textwidth]{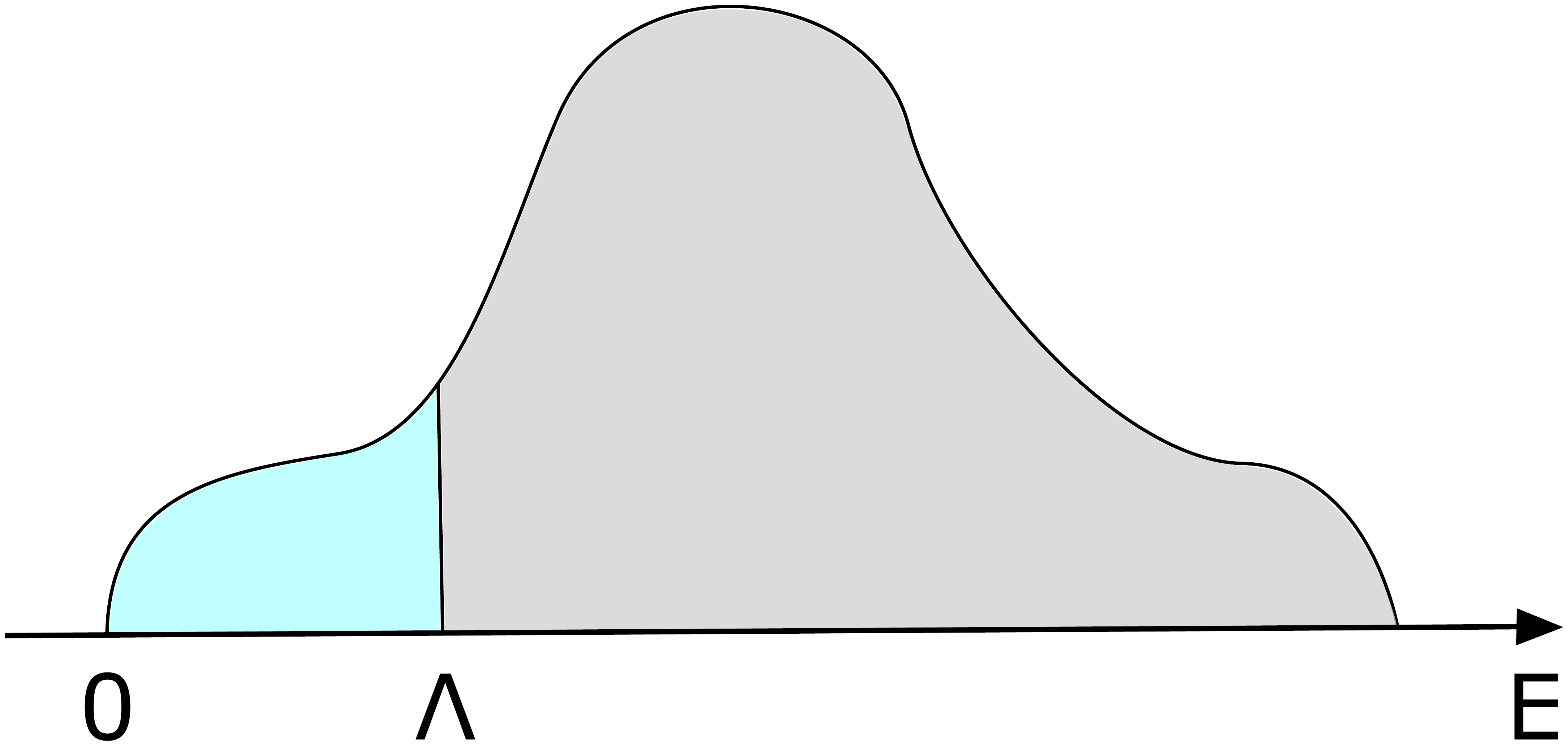}}\label{ensemblespec}
\end{equation}
The blue region represents JT gravity with spectral density \eqref{wiggle}. By definition, our discrete system \eqref{spikes} can be thought of as a single Hamiltonian $M$ of such a matrix ensemble. 
Let us denote its lowest $N$ eigenvalues by $\lambda_1\dots \lambda_N$. 
\\
We expect that the IR behavior of our system \eqref{spikes} is accurately described by a modified matrix ensemble where $N$ eigenvalues are kept fixed to $\lambda_1\dots \lambda_N$. This corresponds to a Dyson gas of charged particles equilibrating in an external potential around $N$ static point charges. These fixed charges repel the charged gas, resulting in a void. We expect the spectral density of this new ensemble to essentially follow the spectrum of the discrete system \eqref{spikes} for $E<\Lambda$ and that of the original ensemble \eqref{ensemblespec} for $E>\Lambda$:
\begin{equation}
\average{\rho(E)} = \quad\raisebox{-9mm}{\includegraphics[width=0.3\textwidth]{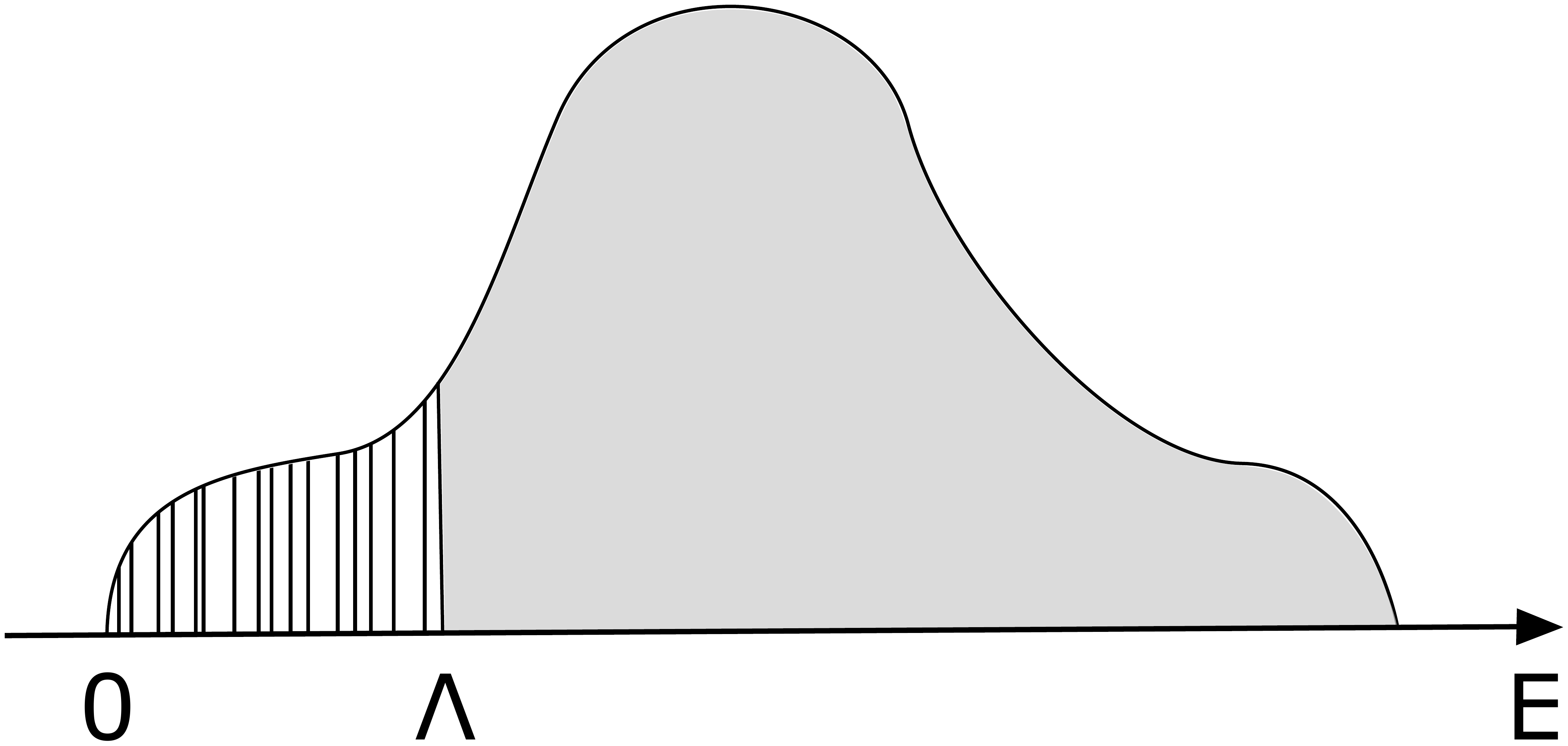}}\label{IRspikes}
\end{equation}
In the remainder of this work, we will make this picture precise and pinpoint its JT gravity interpretation. In particular, we will see that each eigenvalue $\lambda$ corresponds to a fixed energy boundary with label $\lambda$ hovering in the Euclidean bulk. The contour \eqref{dg} in the gravitational path integral is hence over all Riemann surfaces that end on the union of the asymptotic boundaries and on $N$ fixed energy boundaries with labels $\lambda_1\dots \lambda_N$, as shown in \eqref{graphic1level}. This version of JT gravity is able to capture the IR discreteness of \eqref{spikes}. In particular, we will recover the spectral form factor \eqref{plotsff} including erratic oscillations.
\\
In this picture, smooth geometry in the bulk is never in jeopardy: it is provided by our ignorance of the UV part of the system \eqref{spikes}, which corresponds to the $L\gg N$ eigenvalues that remain in the continuum of the matrix integral.
\section{Multi-boundary correlators}\label{sect:preliminary}
This section prepares for section \ref{sect:fixing}, where we will encounter multi-spectral density correlators $\average{\rho(E_1)\dots \rho(E_n)}$. We discuss an efficient way to calculate all significant perturbative and nonperturbative contributions for $e^{S_0}\gg 1$ based on \cite{0408039}.

\subsection{Genus expansion}\label{sect:31}
In JT gravity, it is natural to consider fixed-length boundary conditions, as discussed around \eqref{what}. These correspond to the insertion of macroscopic loop operators in the matrix integral, and are the Laplace transforms of the multi-spectral densities:
\begin{equation}
    Z_\text{JT}(\beta_1\dots \beta_n)=\int_\mathcal{C} d \lambda_1\, e^{-\beta_1\lambda_1}\dots \int_\mathcal{C} d \lambda_n\, e^{-\beta_n\lambda_n}\,\average{\rho(\lambda_1)\dots \rho(\lambda_n)}.\label{ncorrdef}
\end{equation}
This relation is an efficient tool to calculate the perturbative contributions to $\rho(E_1\dots E_n)$ \cite{sss2}. For example, the genus $g$ contribution to the $n$-loop correlation function is:
\begin{align}
\nonumber e^{\chi S_0}\int_0^\infty b_1 d b_1 Z_\text{JT}(\beta_1,b_1)\dots \int_0^\infty b_n d b_n Z_\text{JT}(\beta_2,b_n)\, V_{g,n}(b_1\dots b_n) =\raisebox{-15.5mm}{\includegraphics[width=45mm]{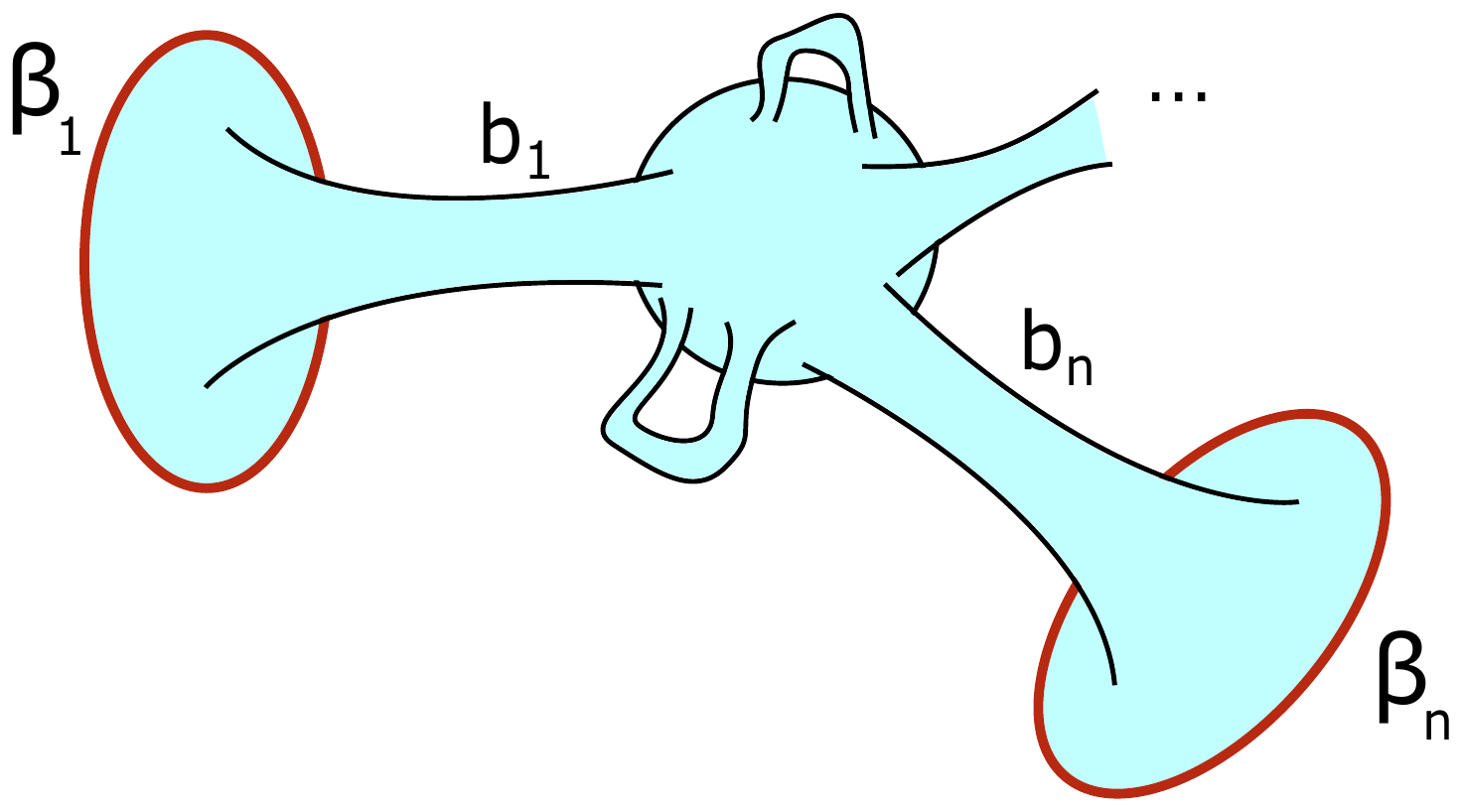}}
\end{align}
Here $V_{g,n}(b_1\dots b_n)$ is the volume of the moduli space of the $n$-holed sphere with $g$ handles. This Weil-Petersson volume is a polynomial in $b_1^2$, $b_2^2$ etc and is easily calculated recursively \cite{mirzakhani}. The single twisted Schwarzian partition function $Z_\text{JT}(\beta,b)$ is just a Gaussian:\footnote{See \cite{Stanford:2017thb,sss2,Mertens:2019tcm} for the evaluation and interpretation of the twisted Schwarzian partition function.} 
\begin{equation}
    Z_\text{JT}(\beta,b)=\frac{e^{-\frac{b^2}{\beta}}}{(\pi\beta)^{1/2}}.\label{sch}
\end{equation}
The Gaussian $b$-integrals yield polynomials in $\beta_1$, $\beta_2$ etc multiplied by $(\beta_1\dots \beta_n)^{1/2}$. Inverse Laplace transforming then gives us the spectral densities, which are polynomials in $1/E_1$, $1/E_2$ etc multiplied by $(E_1\dots E_n)^{-3/2}$. The only exceptions to this polynomial behavior are the disk-and annulus topologies, for which the Weil-Petersson volumes $V_{0,1}(b_1)$ and $V_{0,2}(b_1,b_2)$ are undefined.\footnote{One could take $V_{0,2}(b_1,b_2)=b_1^{-1}\delta(b_1-b_2)$.} The disk density of states is \eqref{diskspec}, and the annulus amplitude is:
\begin{equation}
		\int_0^\infty b db\, Z_\text{JT}(\beta_1,b)\,Z_\text{JT}(\beta_2,b).\label{annuluscalc}
\end{equation}
Its contribution to the two-level spectral density is:
\begin{equation}
    \rho(E_1,E_2) \supset  -\frac{1}{4\pi^2}\frac{E_1+E_2}{\sqrt{E_1}\sqrt{E_2}(E_1-E_2)^2} \approx -\frac{1}{2\pi^2}\frac{1}{(E_1-E_2)^2},\label{annulus}
\end{equation}
where we approximated $\abs{E_1-E_2}\ll 1$. We see that with the exception of the disks and annuli, all the perturbative contributions are small corrections as long as we stay far enough from the spectral edge.\footnote{We should take $E\gg e^{-2S_0/3}$, which ensures the topological suppression prevails over the poles in the spectral densities for small energies, the weakest and hence most important of which is $(E_1\dots E_2)^{-3/2}$. This is confirmed from the exact analysis near the spectral edge in appendix \ref{app:airy} where $E\gg e^{-2S_0/3}$ is the condition for higher loop contributions to the Airy function to become negligible \cite{0408039}.} The annulus contribution itself can become large and comparable to the size of the disk contribution $\rho_0(E_1)\rho_0(E_2)$ for $\abs{E_1-E_2}\sim 1/\rho_0(E)$, the typical eigenvalue spacing in the ensemble. For much larger separations it is negligible. As $\rho_0(E) \sim e^{S_0}$ and we are steering clear of the spectral edge, the only significant contributions from the annulus arise well within the range $\abs{E_1-E_2}\ll 1$. This validates using the second equality in \eqref{annulus} throughout.\\ 
In conclusion, when away from the spectral edge and in the regime $e^{S_0}\gg 1$, all perturbative contributions to the spectral densities are negligible except for those associated with the disk-and annuli topologies.
\\~\\
As the inverse Laplace transforms of fixed length correlators in JT gravity, the spectral densities $\rho(E_1\dots E_n)$ correspond to imposing certain fixed energy boundary condition at the $n$ boundaries of the Riemann surfaces. These boundary conditions are closely related to the FZZT boundary conditions in Liouville theory \cite{0001012}.\footnote{See for example \cite{sss2,paper7}.}

\subsection{Exact answer}\label{sect:32}
On top of the perturbative contributions discussed in the previous subsection, an exact matrix integral analysis reveals nonperturbative contributions to $\rho(E_1\dots E_2)$. We will now discuss an efficient way to calculate all significant contributions in the regime $e^{S_0} \gg 1$, perturbative and non-perturbative, with more details in appendix \ref{app:calculations}.
\\~\\
Brane operators in our matrix ensemble are defined as:
\begin{equation}
        \psi(E)=e^{-\frac{L V(E)}{2}}\prod_{i=1}^L(E-\lambda_i).\label{branedef}
\end{equation} 
We can use this to extract and write the dependence on $\lambda_1$ of the Vandermonde determinant in \eqref{ensemble} as:
\begin{equation}
e^{-L \sum_{i=1}^{L}V(\lambda_i)}\Delta(\lambda_1\dots)=\psi^2(\lambda_1)e^{-L\sum_{i=2}^{L}V(\lambda_i)}\Delta(\lambda_2\dots),\label{brane}
\end{equation}
or more generally:
\begin{equation}
e^{-L \sum_{i=1}^{L}V(\lambda_i)}\Delta(\lambda_1\dots) = \Delta(\lambda_1\dots \lambda_n)\psi^2(\lambda_1)\dots \psi^2(\lambda_n) \,e^{-L\sum_{i=n+1}^{L}V(\lambda_i)}\Delta(\lambda_{n+1}\dots).
\end{equation}
This essentially decomposes the measure of the matrix ensemble \eqref{ensemble}:
\begin{equation}
    d\mu(\lambda_1\dots) = d\lambda_1\dots d\lambda_n\Delta(\lambda_1\dots\lambda_n)\psi^2(\lambda_1)\dots\psi^2(\lambda_n)\,d\mu(\lambda_{n+1}\dots).\label{meassuredecompose}
\end{equation}
For the branes that feature in this formula, the product in \eqref{branedef} is over the $L-n$ remaining eigenvalues.\footnote{Imposing additional symmetries such as time-reversal invariance, changes the matrix ensemble and changes the power of the branes in this formula to 1 or 4, for the GOE or GSE ensembles. We will not study these generalizations here.} This basic formula allows us to extract exact formulas for spectral densities, which we can easily calculate exactly. Let us demonstrate this case-by-case.
\begin{itemize}
    \item \textbf{1 eigenvalue.} We can use the symmetries of the ensemble, and the property \eqref{meassuredecompose} to rewrite \eqref{zb} as (see \eqref{ensemble}):
    \begin{align}
        \nonumber \average{Z(\beta)}&=\frac{L}{\mathcal{Z}_L}\int d\lambda_1\, e^{-\beta \lambda_1} \int d\lambda_2\dots e^{-L V(\lambda_1\dots)} \Delta(\lambda_1\dots)\\
        &=\frac{L\mathcal{Z}_{L-1}}{\mathcal{Z}_L}\int d\lambda_1\, e^{-\beta_1\lambda_1}\, \average{\psi^2(\lambda_1)}_{\LL-1}.\label{zbeta1}
    \end{align}
    From this, we read off the spectral density \cite{sss2}:\footnote{All averaged quantities are $L$ independent for $L\gg 1$.}
    \begin{equation}
        \average{\rho(E)}=\frac{L\mathcal{Z}_{L-1}}{\mathcal{Z}_L}\average{\psi^2(E)}_{\LL-1}.
    \end{equation}
    Both sides of this equality can be calculated independently in JT gravity: in appendix \ref{app:calculations} we calculate the double brane correlator using techniques of \cite{0408039}, and the spectral density was calculated using related techniques but via a different computation in \cite{sss2}. We find:
    \begin{equation}
        \boxed{\average{\rho(E)}=\frac{\average{\psi^2(E)}_{\LL-1}}{2\pi}.}\label{1levelspectral}
    \end{equation}
    Comparison gives us a recursion relation for the matrix integral partition function at large $L$:\footnote{This recursion relation holds for all double-scaled matrix models. It also holds for the CUE ensemble exactly, see e.g. \cite{Gharibyan:2018jrp}.}
    \begin{equation}
        \mathcal{Z}_L\approx 2\pi L \mathcal{Z}_{L-1}.\label{recursive}
    \end{equation}
    This will enable us to eliminate any dependence on $\mathcal{Z}_L$ from the calculations that follow.
    \item \textbf{2 eigenvalues.} The $2$-boundary correlator decomposes as:
    \begin{align}
        \average{Z(\beta_1)Z(\beta_2)}\nonumber  &=\frac{1}{\mathcal{Z}_L}\int d\lambda_1 \dots e^{-LV(\lambda_1\dots)}\Delta(\lambda_1\dots)\sum_{i=1}^L e^{-\beta_1\lambda_i}\sum_{j=1}^L e^{-\beta_2\lambda_j}\nonumber\\
        &=\frac{L}{\mathcal{Z}_L}\int d\lambda_1\, e^{-(\beta_1+\beta_2)\lambda_1}\int d\lambda_2\dots e^{-LV(\lambda_1\dots)}\Delta(\lambda_1\dots)\nonumber\\
        &\quad+\frac{L(L-1)}{\mathcal{Z}_L}\int d\lambda_1\,e^{-\beta_1\lambda_1}\int d\lambda_2\,e^{-\beta_2\lambda_2}\int d\lambda_3\dots e^{-LV(\lambda_1\dots)}\Delta(\lambda_1\dots)\nonumber\\&=\frac{L\mathcal{Z}_{L-1}}{\mathcal{Z}_L}\int d\lambda_1\, e^{-(\beta_1+\beta_2)\lambda_1}\,\average{\psi^2(\lambda_1)}_{\LL-1}\nonumber\\
        &\quad+\frac{L(L-1)\mathcal{Z}_{L-2}}{\mathcal{Z}_L}\int d\lambda_1\,e^{-\beta_1\lambda_1}\int d\lambda_2\,e^{-\beta_2\lambda_2}\,(\lambda_1-\lambda_2)^2\average{\psi^2(\lambda_1)\psi^2(\lambda_2)}_{\LL-2}.\label{zz}
    \end{align}
    Using the recursive formula \eqref{recursive}, we end up with:
    \begin{equation}
        \boxed{\average{\rho(E_1)\rho(E_2)}=\frac{1}{(2\pi)^2}(E_1-E_2)^2\average{\psi^2(E_1)\psi^2(E_2)}_{\LL-2}+\delta(E_1-E_2)\frac{1}{2\pi}\average{\psi^2(E_1)}_{\LL-1}.}\label{2levelspectral}
    \end{equation}
\end{itemize}
These types of formulas are well-known in the random matrix literature. In fact they are referred to simply as the correlation functions \cite{mehta}:\footnote{The constant in formula (6.1.1) of \cite{mehta} is $1/\mathcal{Z}_L$. The average in (6.1.2) generates a factor $\mathcal{Z}_{L-n}$, the recursion relation \eqref{recursive} removes the combinatorial prefactors.}$^,$\footnote{Brane operators in the matrix integral are closely related to exponentiated spacetimes attached to a brane, see appendix \ref{app:calculations}. In this sense, formulas of the type \eqref{1levelspectral} are quite surprising, since they say that a brane-pair correlator actually corresponds to a single (fixed-energy) boundary in gravity.
}
\begin{equation}
    R(E_1\dots E_n)=\frac{1}{(2\pi)^n}\Delta(E_1\dots E_n) \average{\psi^2(E_1)\dots \psi^2(E_n)}_{\LL-n}.\label{rbrane}
\end{equation}
These are smooth functions. The operators in \eqref{rbrane} represent the repulsive force exerted by a set of charges at $E_1\dots E_2$ on the remainder of the Dyson gas. 
\begin{itemize}
    \item \textbf{3 eigenvalues.} An equally easy calculation holds for for the $3$-level spectral density. We find:
    \begin{align}
        \nonumber\average{\rho(E_1)\rho(E_2)\rho(E_3)}=&R(E_1,E_2,E_3)+\delta(E_1-E_2)R(E_1,E_3)+\delta(E_1-E_3)R(E_2,E_3)\\
        &+\delta(E_2-E_3)R(E_1,E_2)+\delta(E_1-E_2)\delta(E_2-E_3)R(E_1).\label{3}
    \end{align}
    This is readily generalized to any number of boundaries.
\end{itemize}
The delta-functions that appear in these expressions of this kind are contact terms. Whereas a geometric interpretation of the correlation functions $R(E_1\dots E_n)$ is obvious from the discussion of section \ref{sect:31}, the interpretation of these terms is somewhat obscure. We will come back to this in the concluding section \ref{sect:concl}.
\\~\\
It is convenient to extract from the correlation functions $R(E_1\dots E_n)$ the fully connected contribution $T(E_1, \hdots E_n)$ known as the cluster function. The remaining disconnected pieces are then products of cluster functions at lower values of $n$. For example \cite{mehta}:\footnote{The minus signs are convention \cite{mehta}.}
\begin{align}
R(E) =&\, T(E), \nonumber \\
R(E_1,E_2) =& -T(E_1,E_2) + T(E_1)T(E_2), \nonumber\\
R(E_1,E_2,E_3) =&\, T(E_1,E_2,E_3) - T(E_1)T(E_2,E_3) -T(E_2)T(E_1,E_3)\nonumber\\
&-T(E_3)T(E_1,E_2)+ T(E_1)T(E_2)T(E_3).
\label{ancluster}
\end{align}
Following the logic around \eqref{conngeom}, one immediately deduces that the clusters $T(E_1,\hdots E_n)$ correspond to the nonperturbative completion of the gravitational genus expansion starting with the $n$-holed sphere.\footnote{The precise version of this statements follows from the decomposition of the correlators $\average{\rho(E_1)\dots\rho(E_n)}$ into cluster functions, including contact terms. The resulting cluster functions correspond precisely to the nonperturbative completion of the $n$-holed sphere genus expansion, which will generate the same contact terms. For example, the three-holed sphere with all corrections gives:
\begin{align}
    \nonumber \average{\rho(E_1)\rho(E_2)\rho(E_2)}^\text{conn}=&T(E_1,E_2,E_3)-\delta(E_1-E_2)T(E_1,E_3)-\delta(E_1-E_3)T(E_1,E_2)\\&-\delta(E_2-E_3)T(E_1,E_3)+\delta(E_1-E_2)\delta(E_1-E_3)T(E_1).
\end{align}
This follows directly from \eqref{3}, but also follows intuitively from the discussion on merging boundaries in \ref{sect:concl}.
} The cluster functions have the property that they vanish when the spacing of two of its arguments is large compared to the average eigenvalue spacing. This means the only significant contributions of the cluster functions to the correlation functions $R(E_1\dots E_n)$ are when $\abs{E_i-E_j}\ll 1$ for all energies in a cluster.
\\
The perturbative disk-and annuli contributions discussed in section \ref{sect:31} are part of the terms $T(E_i)$ respectively $T(E_i,E_j)$ that contribute a generic correlator $R(E_1\dots E_n)$. As mentioned earlier, these are the only significant perturbative contributions to $R(E_1\dots E_n)$ away from the spectral edge. 
\\~\\
An exact calculation of the correlators $R(E_1\dots E_n)$ in JT gravity reveals a set of non-perturbative contributions similar to those in \eqref{nptwo}. An efficient way to calculate these exactly in JT gravity is via formula \eqref{rbrane}. We do so in appendix \ref{app:calculations} in detail for $R(E_1)$ and $R(E_1,E_2)$ and discuss certain aspects of the calculation for $R(E_1,E_2,E_3)$. The general trend is the appearance of significant non-perturbative contributions to $R(E_1\dots E_n)$ of the type:
\begin{equation}
    \frac{\exp(\pm i\pi \int_{E_i}^{E_j} d M\,\rho(M))}{(E_i-E_j)}.\label{contri}
\end{equation}
It is convenient to extract from this the cluster functions, which as explained above can be evaluated for $\abs{E_i-E_j}\ll 1$. We find:
\begin{align}
    \nonumber T(E)&=\rho(E)\\
    T(E_1,E_2)&=\rho(E_1)\rho(E_2)\,\sinc^2\rho(E_1)(E_1-E_2)=S(E_1,E_2)^2.\label{cluster}
\end{align}
The sine kernel $S(E_1,E_2)$ also appears in higher clusters, for example:
\begin{equation}
    T(E_1,E_2,E_3)=2\,S(E_1,E_2)\,S(E_2,E_3)\,S(E_3,E_1).
\end{equation}
This is very unsurprising. It is a widely held conjecture \cite{mehta} for any Hermitian matrix ensemble that cluster functions are exactly equal to the universal GUE cluster functions when its arguments are close together $\abs{E_i-E_j}\ll 1$. The latter are known exactly \cite{mehta} and feature only the sine kernel. In the brane calculations these arise due to the contributions of the type \eqref{contri}. The calculations of appendix \ref{app:calculations} merely reassure us that this conjecture is true in JT gravity. We are then free to ship in the GUE clusters to calculate $\average{\rho(E_1)\dots \rho(E_n)}$ in JT gravity.
\section{Fixing eigenvalues or introducing boundaries}\label{sect:fixing}
In this section we investigate a matrix ensemble with a series of eigenvalues fixed to consecutive ones of \eqref{spikes}, and specify the integration space in the JT gravity path integral over metrics \eqref{dg} associated to this ensemble. The specific contour follows from formula \eqref{rbrane} combined with \eqref{meassuredecompose}: each fixed eigenvalue corresponds to an additional fixed-energy boundary in the bulk on which Riemann surfaces can end.
\\
A matrix ensemble with $n$ eigenvalues fixed to $\lambda_1 \hdots \lambda_n$ (assumed all different) is obtained from the original ensemble \eqref{ensemble} by including appropriate deltas in the measure:\footnote{Here $d\mu(\kappa_1\dots \kappa_L)$ is the measure of \eqref{ensemble}.}
\begin{equation}
    d\mu(\kappa_1\dots \kappa_L)\prod_{i=1}^L \delta(\kappa_i-\lambda_i).
\end{equation}
The partition function replacing \eqref{ensemble} is then:
\begin{align}
\nonumber\mathcal{Z}_{L,\lambda_1\dots \lambda_n} &= \int d\lambda_{n+1}\dots d\lambda_L\,e^{-L V(\lambda_1\dots)}\,\Delta(\lambda_1\dots)=\mathcal{Z}_{L-n} \Delta(\lambda_1\dots\lambda_n)\average{\psi^2(\lambda_1)\dots \psi^2(\lambda_n)}_{\LL-n} \nonumber \\
&= (2\pi)^n \mathcal{Z}_{L-n}\, \average{\rho(\lambda_1)\dots \rho(\lambda_n)}_{\LL}.\label{thermalensemble}
\end{align}
Here we used \eqref{meassuredecompose} and the generalization of \eqref{3} to $n$ boundaries. Notice that the contact terms vanish because the eigenvalues of \eqref{spikes} are all different. Perturbatively, this partition function is counting Riemann surfaces of the type:
\begin{equation}
   \average{\rho(\lambda_1)\dots \rho(\lambda_n)} \supset \quad \raisebox{-18.5mm}{\includegraphics[width=15mm]{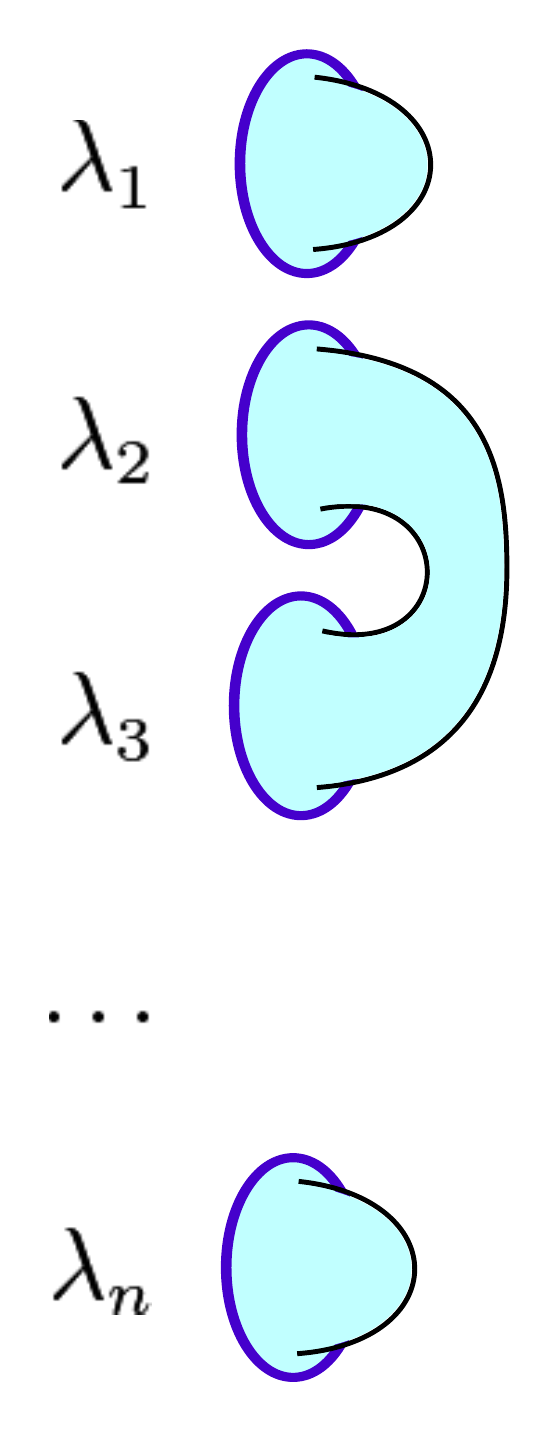}}
\end{equation}
where $n$ eigenbrane boundaries are present, but no asymptotic boundary insertions.
\subsection{Delta spikes and a void}
The expectation value of the spectral density $\rho(E)=\sum_{i=1}^L \delta(E-\lambda_i)$ in the new ensemble \eqref{thermalensemble} is by definition:
\begin{equation}
\average{\rho(E)}_{\lambda_1\dots\lambda_n}= \frac{1}{\mathcal{Z}_{L,\lambda_1\dots \lambda_n}}\int d\lambda_{n+1}\dots d\lambda_L\,\rho(E) e^{-L V(\lambda_1\dots)}\,\Delta(\lambda_1\dots).
\end{equation}
We immediately obtain:
\begin{equation}
\boxed{\average{\rho(E)}_{\lambda_1\dots\lambda_n}= \frac{\average{\rho(E)\rho(\lambda_1)\dots\rho(\lambda_n)}_{\LL}}{\average{\rho(\lambda_1)\dots\rho(\lambda_n)}_{\LL}}.}\label{perturbativerho}
\end{equation}
This is a conditional probability. As announced, this corresponds to a version of JT gravity where each fixed eigenvalue of the matrix integral translates into the introduction of a fixed-energy boundary on which Riemann surfaces in the path integral are to end. As explained before, in the genus expansion disks and annuli dominate the regime of interest. From \eqref{perturbativerho} we read off the type of geometries contributing significantly to the JT gravity path integral:
\begin{equation}
   \average{\rho(\lambda_1)\dots\rho(\lambda_n)} \average{\rho(E)}_{\lambda_1\dots\lambda_n}\supset \quad \raisebox{-18mm}{\includegraphics[width=35mm]{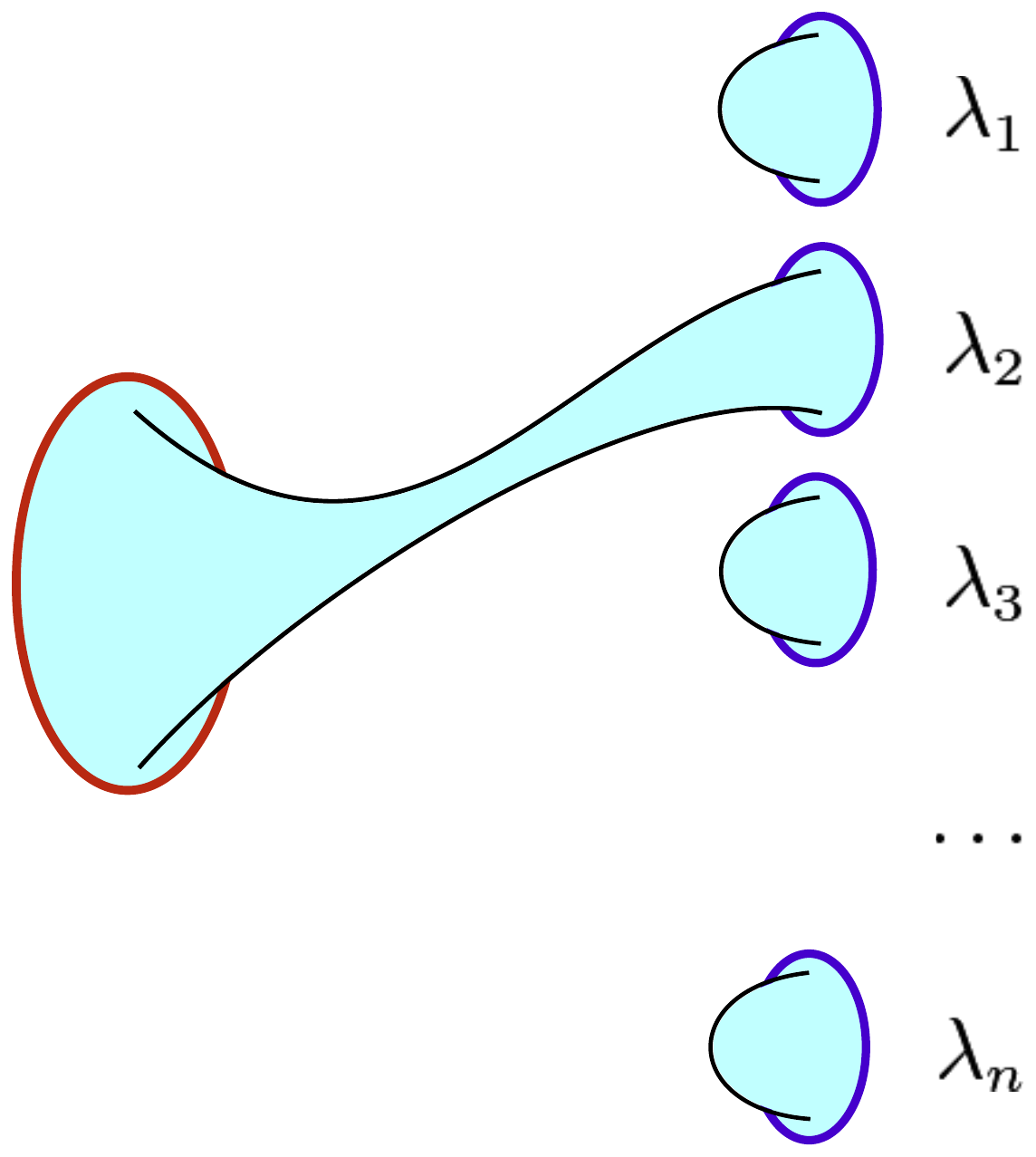}}\label{graphic1level}
\end{equation}
There are also multi-annulus configurations where eigenvalue boundaries connect to other eigenvalue boundaries. Three-holed spheres and handle-body geometries contribute, but not significantly. 
\\
Using a generalization of formula \eqref{3} we can rewrite \eqref{perturbativerho} as:
\begin{equation}
    \average{\rho(E)}_{\lambda_1\dots\lambda_n} = \sum_{i=1}^{n} \delta(E-\lambda_i) + \frac{R(E,\lambda_1\dots\lambda_n)}{R(\lambda_1\dots\lambda_n)}.\label{densityfree}
\end{equation}
At this point our discussion of the previous section comes into play: we can immediately write down the exact answer for a given $n$ using the cluster functions \eqref{cluster} etc. 
\\~\\
As a consistency check on the normalization, we can take the integral over $E$ of \eqref{densityfree} using the result of Appendix \ref{app:add}:
   \begin{equation}
        \int_\mathcal{C}d\lambda \frac{R(\lambda,\lambda_1\dots \lambda_n)}{R(\lambda_1\dots \lambda_n)}=L-n.\label{toproof}
    \end{equation}
We see that the number of eigenvalues in the smooth continuum is exactly down by $n$ as compared to $\rho(E)$, and these eigenvalues are accounted for by the delta functions.
\\
As discussed in section \ref{sect:preliminary}, the contributions from the annuli connecting the asymptotic boundary to the eigenvalue boundaries is negligible when $\abs{E-\lambda_i}\gg 1/\rho(E)$, and the same holds for all nonperturbative contributions. Therefore, all effects due to the fixed eigenvalues are short-ranged and one has:\footnote{This corresponds to the intuition of section \ref{sect:213} that far enough from the fixed charges we can't distinguish them from the scenario where the charged gas would fill in this space.}
\begin{equation}
    \average{\rho(E)}_{\lambda_1\dots \lambda_n}\approx \average{\rho(E)},\quad \abs{E-\lambda_i}\gg 1/\rho(E).
\end{equation}
To fully understand the physics in the exact formula \eqref{densityfree}, let us do a small case-by-case study.\footnote{The eigenvalues used to generate these plots are the same as those used in the plot of \eqref{plotsff}. These are exact plots, not cartoons.}
\begin{itemize}
    \item \textbf{1 eigenvalue.} We have from \eqref{densityfree}:
    \begin{align}
        \average{\rho(E)}_{\lambda} &=\delta(E-\lambda)+\rho(E)(1-\sinc^2 \pi \rho(\lambda) (E-\lambda)).\label{rho1cont}
    \end{align}
    Close to the fixed eigenvalue this looks like:
    \begin{align}
        \average{\rho(E)}_\lambda &=\quad\raisebox{-10mm}{\includegraphics[width=50mm]{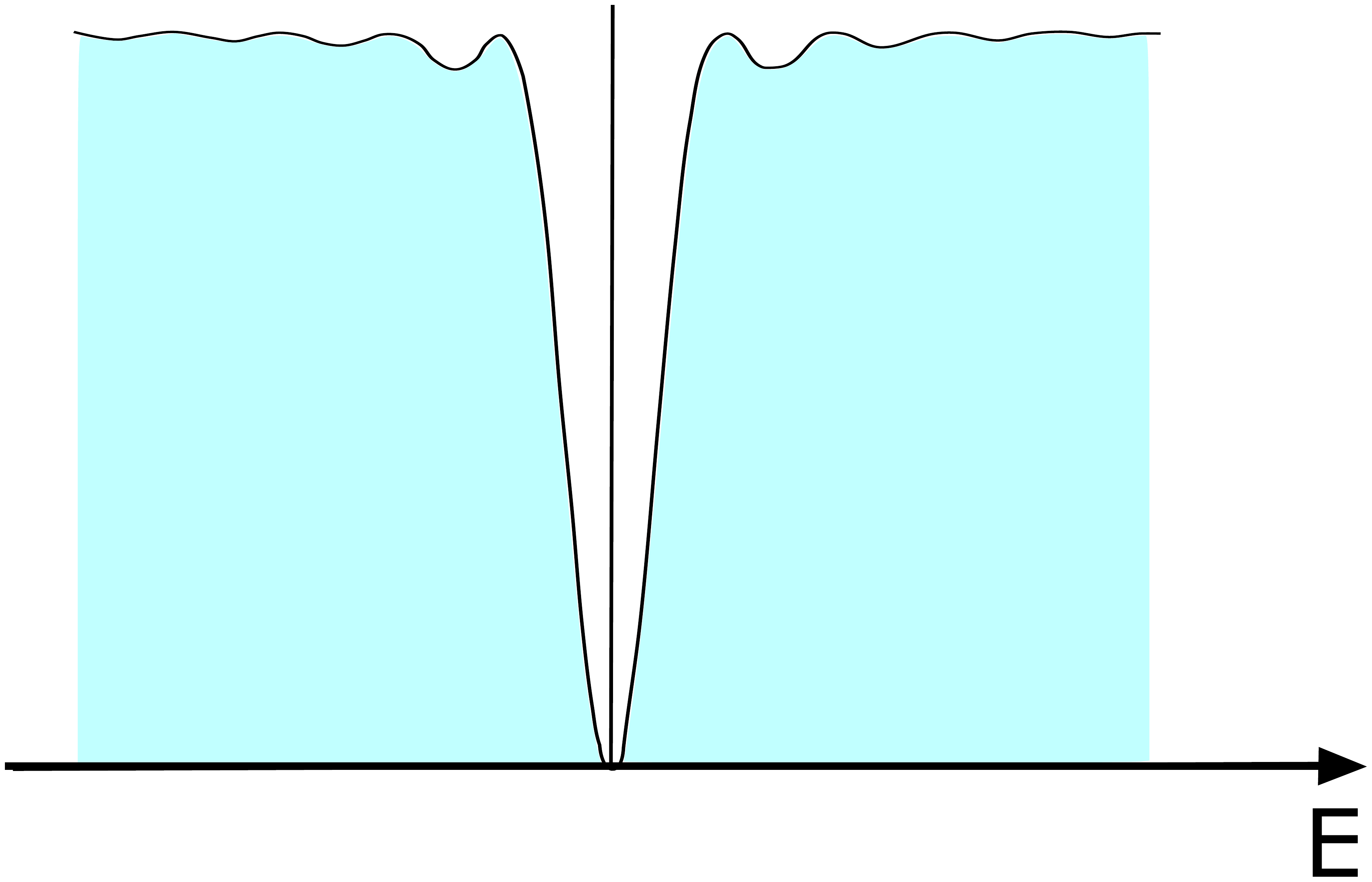}}\label{1fixed}
    \end{align}
    This exhibits eigenvalue repulsion: the fixed charge repels the particles of the gas, as modeled here by the Vandermonde factor $(E-\lambda)^2$.
    \item \textbf{2 eigenvalues.} Using the GUE cluster functions \eqref{cluster} in \eqref{densityfree}, we find a less elegant answer for the case of two fixed eigenvalues.\footnote{
    \begin{align}
        \nonumber \rho(E)_{\lambda_1,\lambda_2}&=\delta(E-\lambda_1)+\delta(E-\lambda_2)+\rho(E)-\rho(E)\frac{\sinc^2 \pi \rho(E)(E-\lambda_1)+\sinc^2 \pi \rho(E)(E-\lambda_2)}{1-\sinc^2 \pi \rho(E) (\lambda_1-\lambda_2)}\\
        &\quad - \rho(E)\frac{2 \sinc \pi \rho(E)(E-\lambda_1)\sinc \pi \rho(E)(E-\lambda_2)\sinc \pi \rho(E)(\lambda_1-\lambda_2)}{1-\sinc^2 \pi \rho(E)(\lambda_1-\lambda_2)}.\label{rho2}
    \end{align}} A plot close to the fixed eigenvalues is much more intuitive:
    \begin{equation}
        \average{\rho(E)}_{\lambda_1,\lambda_2}=\quad \raisebox{-10mm}{\includegraphics[width=50mm]{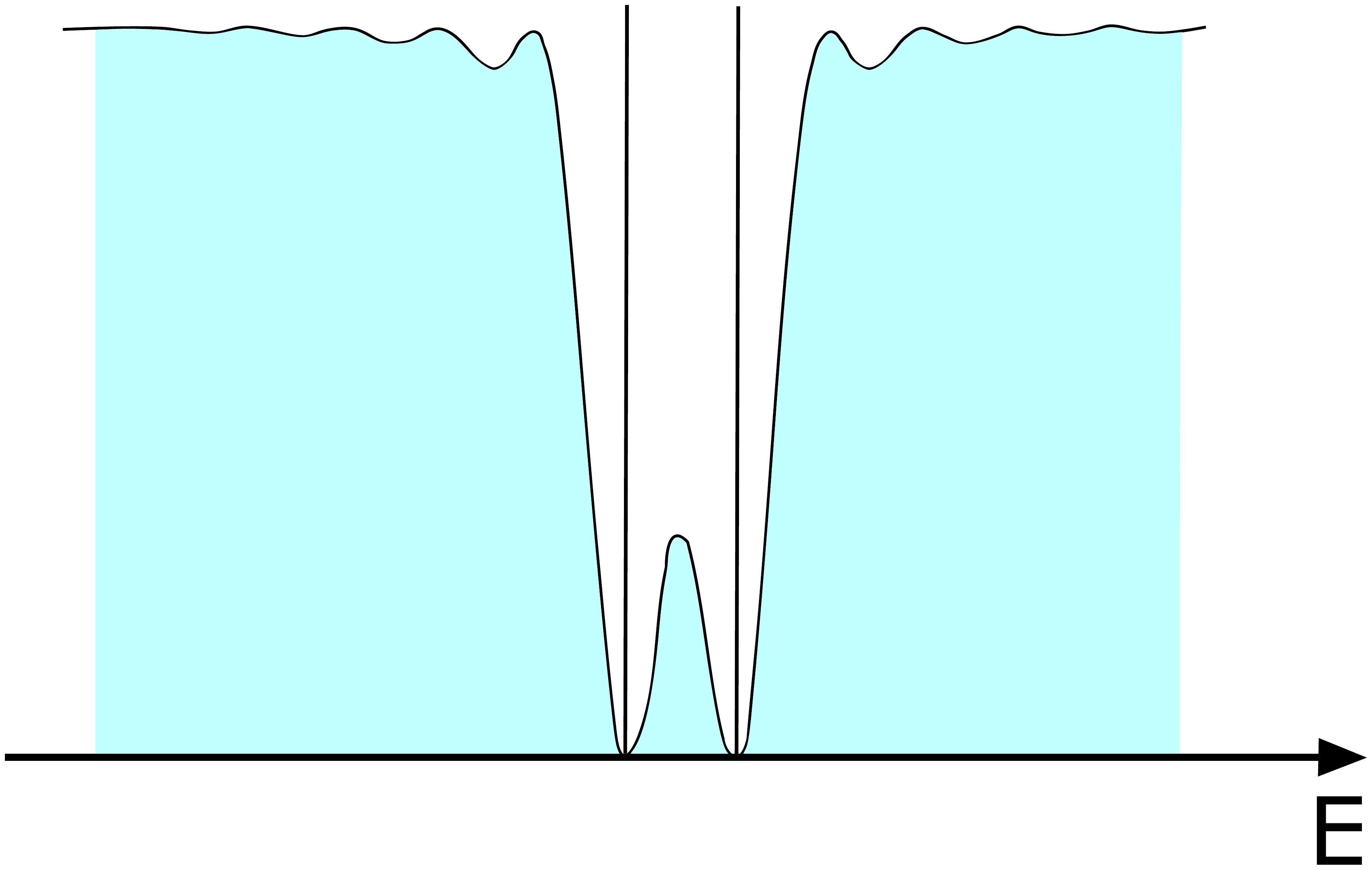}}
    \end{equation}
    There is a relatively low probability for another eigenvalue to be found in between $\lambda_1$ and $\lambda_2$, provided they are close enough. \\
	In general we can think of the initial coarse-grained density as a low-frequency approximation to the series of delta-functions. The maximal frequency here is the typical eigenvalue spacing $1/\rho(E)$. We see therefore manifestly that we are not changing any early-time $t\ll \rho(E)$ physics by fixing eigenvalues.
    \item \textbf{A bin of eigenvalues.} It is not hard to plot \eqref{densityfree} exactly for an increasing number of consecutive eigenvalues of \eqref{spikes} in some region. For example, for $n=8$ we find:
    \begin{equation}
        \average{\rho(E)}_{\lambda_1\dots\lambda_8}=\quad \raisebox{-10mm}{\includegraphics[width=50mm]{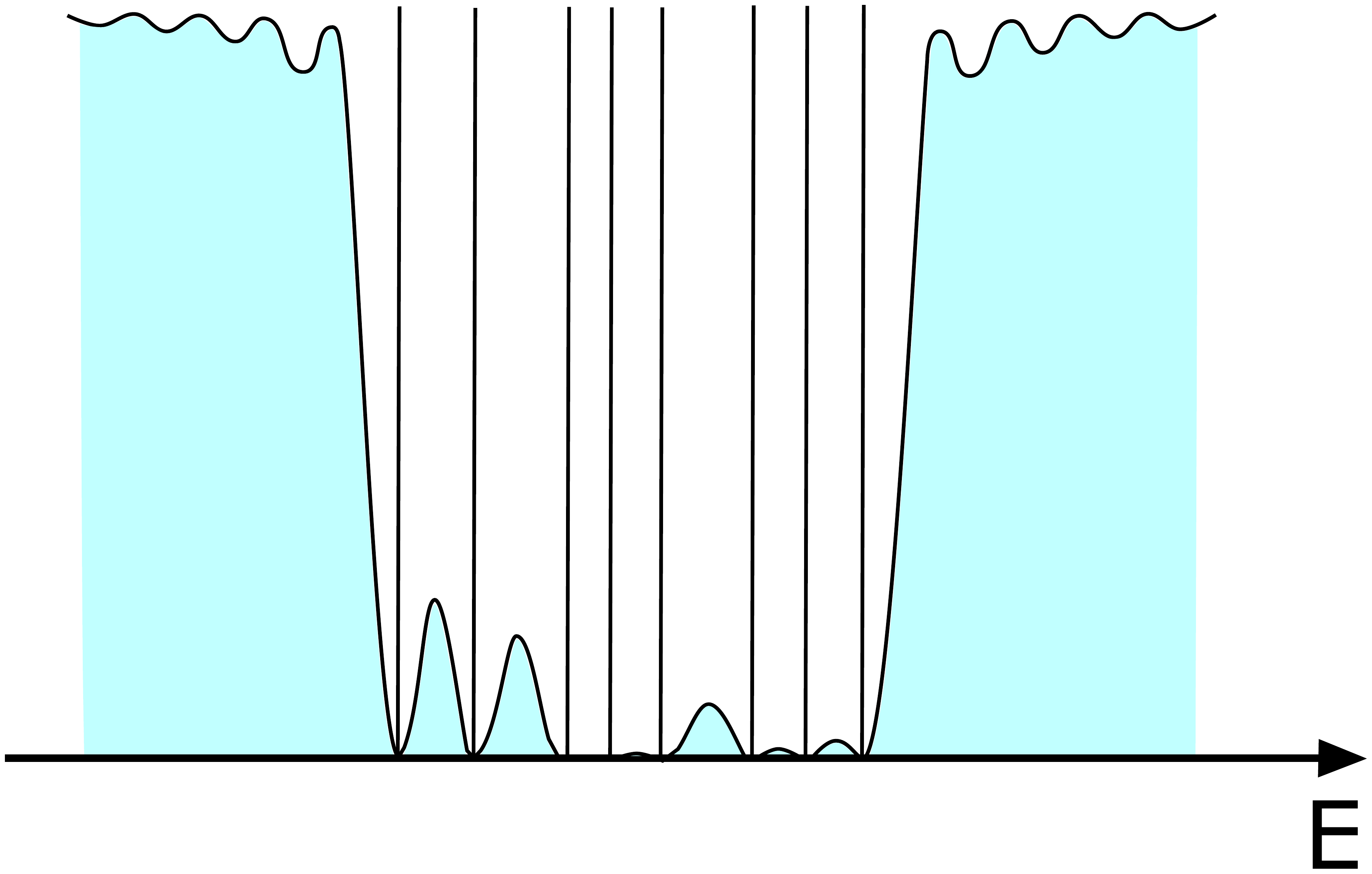}}\label{8spectral}
    \end{equation}
    We are starting to see the features claimed in formula \eqref{IRspikes}. Firstly, fixing a large number of consecutive eigenvalues will create to good approximation a void in the continuum spectral density in the interval $\mathcal{I}$ where the eigenvalues are situated:
    \begin{equation}
        \average{\rho(E)}_{\lambda_1\dots\lambda_n} \approx \sum_{i=1}^n \delta(E-\lambda_i),\quad E\in \mathcal{I}.\label{discretrho}
    \end{equation}
    Secondly, the effect is not felt far outside of $\mathcal{I}$: the effect dies out over a range $ \sim 1/\rho(E)\ll 1$. 
\end{itemize}
As mentioned before, in the region closer to the spectral edge, where the spectral density $\rho(E)$ changes rapidly, we can no longer trust the sine-kernel type GUE cluster functions \eqref{cluster}. Fortunately, in that region, we have available the exact results of the Airy model. Using the method of \cite{0408039} to calculate brane correlators, it is straightforward though slightly tedious to recover the known Airy cluster functions.
\\
We do so in appendix \ref{app:airy} for the case $T(E_1,E_2)$ and recover the Airy kernel \cite{mehta}. We then study the spectral density with one fixed eigenvalue close to the spectral edge. The behavior is very similar to that of \eqref{1fixed}. It would be straightforward to extend this to multiple fixed eigenvalues, but we will refrain from doing so.
\\~\\
All this points in the direction of the picture \eqref{IRspikes}: by inserting the $1\ll N\ll \Lambda$ fixed energy boundaries discussed in section \ref{sect:213} in JT gravity, we get a version of JT gravity with a spectral density that is essentially completely discretized in the region $E<\Lambda$, matching that of the abstract discrete system \eqref{spikes}.
\\
We note that it is possible that the calculations of the brane correlators presented in appendix \ref{app:calculations} are more subtle when $n \sim e^{S_0}$. In particular, the limit $e^{S_0}\gg 1$ used in \cite{0408039} to obtain the semiclassical brane correlators could be more subtle. It would be valuable to understand if this happens, and how the technical calculation is modified. There is no reason though to expect any qualitative deviations from the picture \eqref{IRspikes} and our conclusions. In particular we expect no sizeable modification of the correlation function $R(E_1\dots E_n)$ away from the GUE answer.
\subsection{Erratic oscillations}\label{sect:erratic}
To stack up the claim that introducing these eigenbranes in JT gravity allows one to capture the $E<\Lambda$ features of the discrete system with spectrum \eqref{spikes}, we would like to reproduce the local spectral form factor \eqref{plotsff} from a JT gravity calculation. For this we will investigate $\average{\rho(E_1)\rho(E_2)}_{\lambda_1\dots \lambda_n}$, with emphasis on the terms that contribute to the plateau region $t> 2\pi \rho(E)$.
\\~\\
Using the ensemble with $n$ fixed eigenvalues \eqref{thermalensemble}, one immediately writes down:
\begin{equation}
    \boxed{\average{\rho(E_1)\rho(E_2)}_{\lambda_1\dots\lambda_n}=\frac{\average{\rho(E_1)\rho(E_2)\rho(\lambda_1)\dots\rho(\lambda_n)}}{\average{\rho(\lambda_1)\dots\rho(\lambda_n)}}.}\label{perturbativetwo}
\end{equation}
Geometrically, we are calculating the correlator of two fixed-energy boundaries in a version of JT gravity that has $n$ fixed-energy boundaries hovering in the bulk. The only significant perturbative contributions are due to the disk and annuli, for example:
\begin{align}
    \average{\rho(\lambda_1)\dots \rho(\lambda_2)}\average{\rho(E_1)\rho(E_2)}_{\lambda_1\dots\lambda_n}\supset\, \raisebox{-19mm}{\includegraphics[width=35mm]{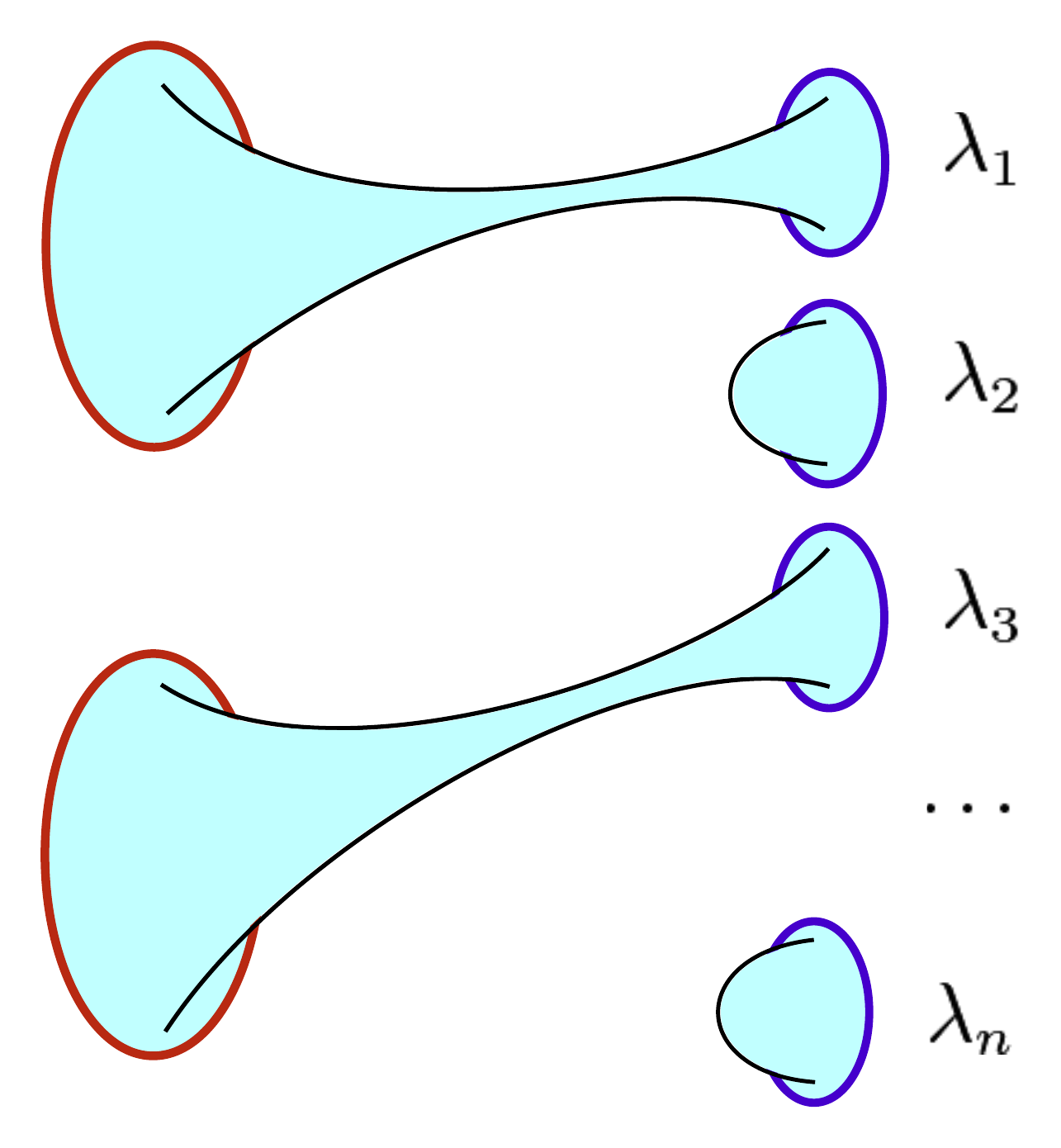}} \quad ,\,\, \raisebox{-19mm}{\includegraphics[width=35mm]{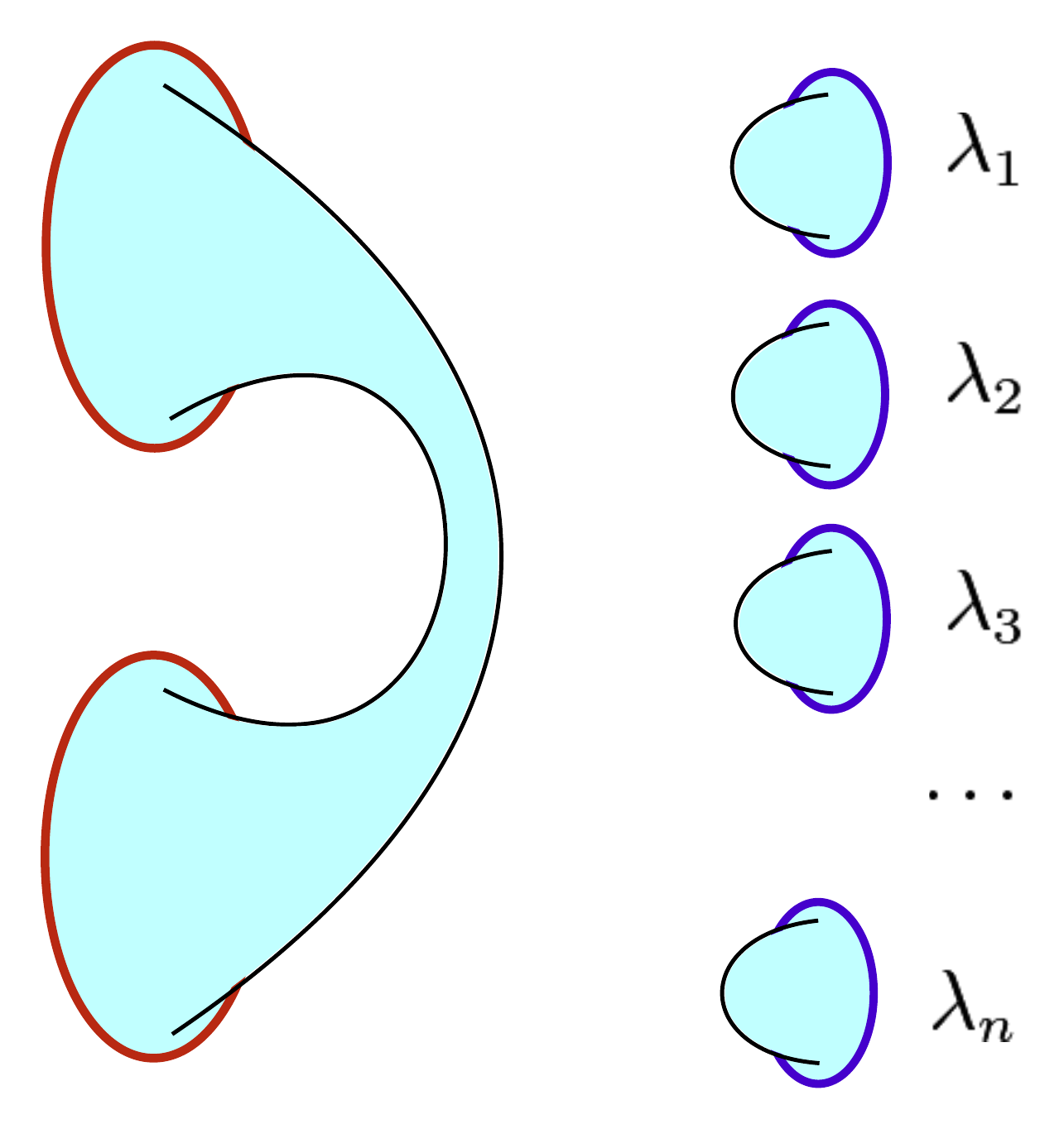}}\label{graphic2level}
\end{align}
As in \eqref{graphic1level} the eigenvalue boundaries that don't connect to the asymptotic boundaries don't need to be capped off by disks, there can be annuli between them. 
\\
Using the exact formulas for the multi-spectral densities discussed in section \ref{sect:preliminary}, we obtain:
\begin{align}
    \average{\rho(E_1)\rho(E_2)}_{\lambda_1\dots\lambda_n}&=\sum_{i=1}^n \delta(E_1-\lambda_i)\sum_{j=1}^n \delta(E_2-\lambda_j)\nonumber\\
    &+\frac{R(E_1,\lambda_1\dots\lambda_n)}{R(\lambda_1\dots\lambda_n)}\sum_{j=1}^n\delta(E_2-\lambda_j)+\frac{R(E_2,\lambda_1\dots\lambda_n)}{R(\lambda_1\dots\lambda_n)}\sum_{i=1}^n\delta(E_1-\lambda_i)\nonumber\\
    &+\delta(E_1-E_2)\frac{R(E_1,\lambda_1\dots\lambda_n)}{R(\lambda_1\dots\lambda_n)}+\frac{R(E_1,E_2,\lambda_1\dots\lambda_n)}{R(\lambda_1\dots\lambda_n)}.\label{twolevelfree}
\end{align}
Again, using the JT spectral density and the GUE cluster functions it is easy to calculate and plot this recursively for increasing $n$. Numerically investigating the continuous contributions to \eqref{twolevelfree} it quickly becomes obvious that if we fix a large number of eigenvalues of \eqref{spikes}, then in the region $\mathcal{I}$ where the eigenvalues are positioned, to good approximation:\footnote{In particular, much like the depletion of the continuum spectral density in for example \eqref{8spectral}, one observes that well within the bulk of $\mathcal{I}\times \mathcal{I}$, the final term in \eqref{twolevelfree} can be made arbitrarily small by increasing $n$.
}
\begin{equation}
    \average{\rho(E_1)\rho(E_2)}_{\lambda_1\dots\lambda_n} \approx \sum_{i=1}^n \delta(E_1-\lambda_i)\sum_{j=1}^n \delta(E_2-\lambda_j),\quad E_1, E_2 \in \mathcal{I}.\label{discrete}
\end{equation}
If we take the region $\mathcal{I}$ large enough such that $\abs{\text{bin}(E)}\ll \abs{\mathcal{I}}$ then we trivially recover the discrete version of the local spectral form factor \eqref{tofind} in JT gravity, including all erratic oscillations in \eqref{plotsff}.
\\
We would like to understand in a bit more detail the approach of the local spectral form factor to this erratic behavior though. Let us focus on the plateau region $t>2\pi \rho(E)$ and take only a few fixed eigenvalues.\footnote{An analytic analysis of the plateau region is simpler than that of the ramp region.} In the averaged version of JT gravity, the plateau behavior is only due to the first term in \eqref{twolevel0}:
\begin{equation}
    \average{\rho(E_1)\rho(E_2)}^\text{plateau}=\delta(E_1-E_2) \rho(E_1).\label{averaged}
\end{equation}
In appendix \ref{app:details} we point out that only the first and penultimate contributions to \eqref{twolevelfree} are relevant for the spectral form factor at $t>2\pi\rho(E)$:
\begin{align}
    \average{\rho(E_1)\rho(E_2)}_{\lambda_1\dots\lambda_n}^\text{plateau}&=\delta(E_1-E_2)\frac{R(E_1,\lambda_1\dots\lambda_n)}{R(\lambda_1\dots\lambda_n)}+\sum_{i,j}^n\delta(E_1-\lambda_i)\delta(E_2-\lambda_j) \nonumber \\
		&= \delta(E_1-E_2)\average{\rho(E_1)}_{\lambda_1\dots\lambda_n} + \sum_{i\neq j}^n \delta(E_1-\lambda_i)\delta(E_2-\lambda_j).
\end{align}
This formula nicely interpolates between the averaged variant \eqref{averaged} and the discretized variant \eqref{discrete}. The first term contributes a constant plateau of height $N$.\footnote{This is a variant of \eqref{toproof} where we take the fixed eigenvalues sufficiently deep in the bin, such that the tails extending outside the bin are negligible. The continuum contributes $N-n$ and the deltas give $n$.} The second term generates ever more erratic oscillations for increasing number of eigenvalues:
\begin{equation}
    \average{S_E(t)}_{\lambda_1\dots \lambda_n} = N + \sum_{i\neq j}^n \cos t(\lambda_i-\lambda_j).
\end{equation}
For $n=0$ this is the usual random matrix theory answer, for $n=N$ we recover the discrete answer \eqref{tofind}.
\subsection{Disconnection}
The connected part of the two level spectral density is defined as:
\begin{equation}
    \average{\rho(E_1)\rho(E_2)}_{\lambda_1\dots\lambda_n}^\text{conn}=\average{\rho(E_1)\rho(E_2)}_{\lambda_1\dots\lambda_n}-\average{\rho(E_1)}_{\lambda_1\dots\lambda_n}\average{\rho(E_2)}_{\lambda_1\dots\lambda_n}.
\end{equation}
From \eqref{discrete} and \eqref{discretrho}, we get to good approximation:
\begin{equation}
\label{conncorr}
    \average{\rho(E_1)\rho(E_2)}_{\lambda_1\dots\lambda_n}^\text{conn} \approx 0,\quad E_1, E_2 \in \mathcal{I}.
\end{equation}
This is trivial for a discrete system, but it entails a nontrivial equality in bulk gravity. 
\\
To appreciate this, consider the geometries that contribute to the two level spectral density $\average{\rho(\lambda_1)\dots \rho(\lambda_2)}\average{\rho(E_1)\rho(E_2)}_{\lambda_1\dots\lambda_n}$ and compare this to the geometries that contribute to $\average{\rho(\lambda_1)\dots \rho(\lambda_2)}\average{\rho(E_1)}_{\lambda_1\dots\lambda_n}\average{\rho(\lambda_1)\dots \rho(\lambda_2)}\average{\rho(E_2)}_{\lambda_1\dots\lambda_n}$. If we strip off geometries that contribute to both, we end up in the former with connected geometries such as the annulus between the two asymptotic boundaries. In the latter we are left with configurations where the boundaries are indirectly connected via matching pairs of eigenbranes. The sum of what remains in either quantity is non-zero. We can calculate the exact answer for each quantity independently for increasing $n$ using the techniques of section \ref{sect:32}. It turns out that these quantities match for $ E_1, E_2 \in \mathcal{I}$. This proves the following geometric property:
\begin{equation}
\label{connect1}
		  \sum_{\text{connected}} \,  \raisebox{-18.5mm}{\includegraphics[width=43mm]{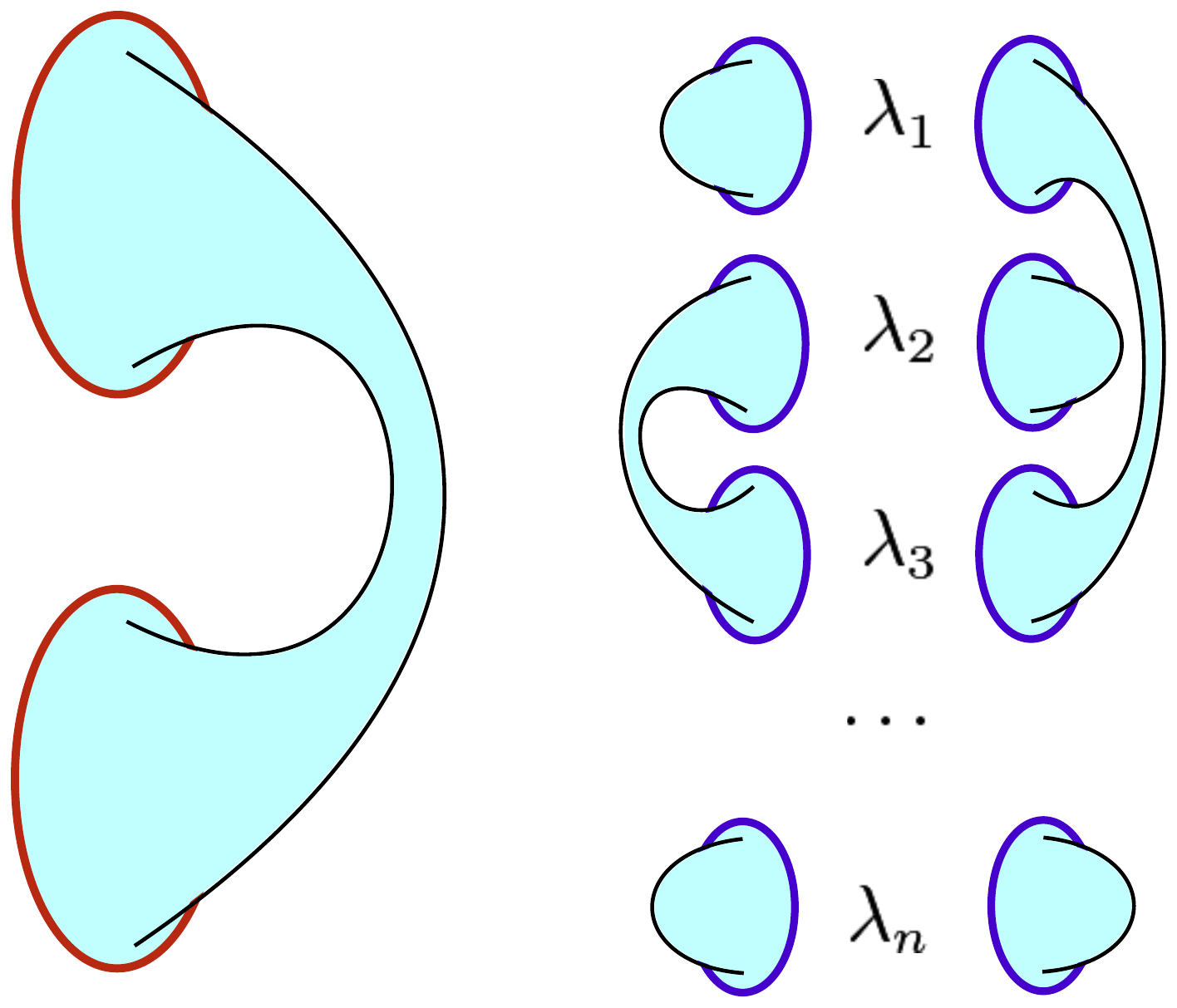}}\, =\,\sum_{\text{connected}} \, \raisebox{-20mm}{\includegraphics[width=67mm]{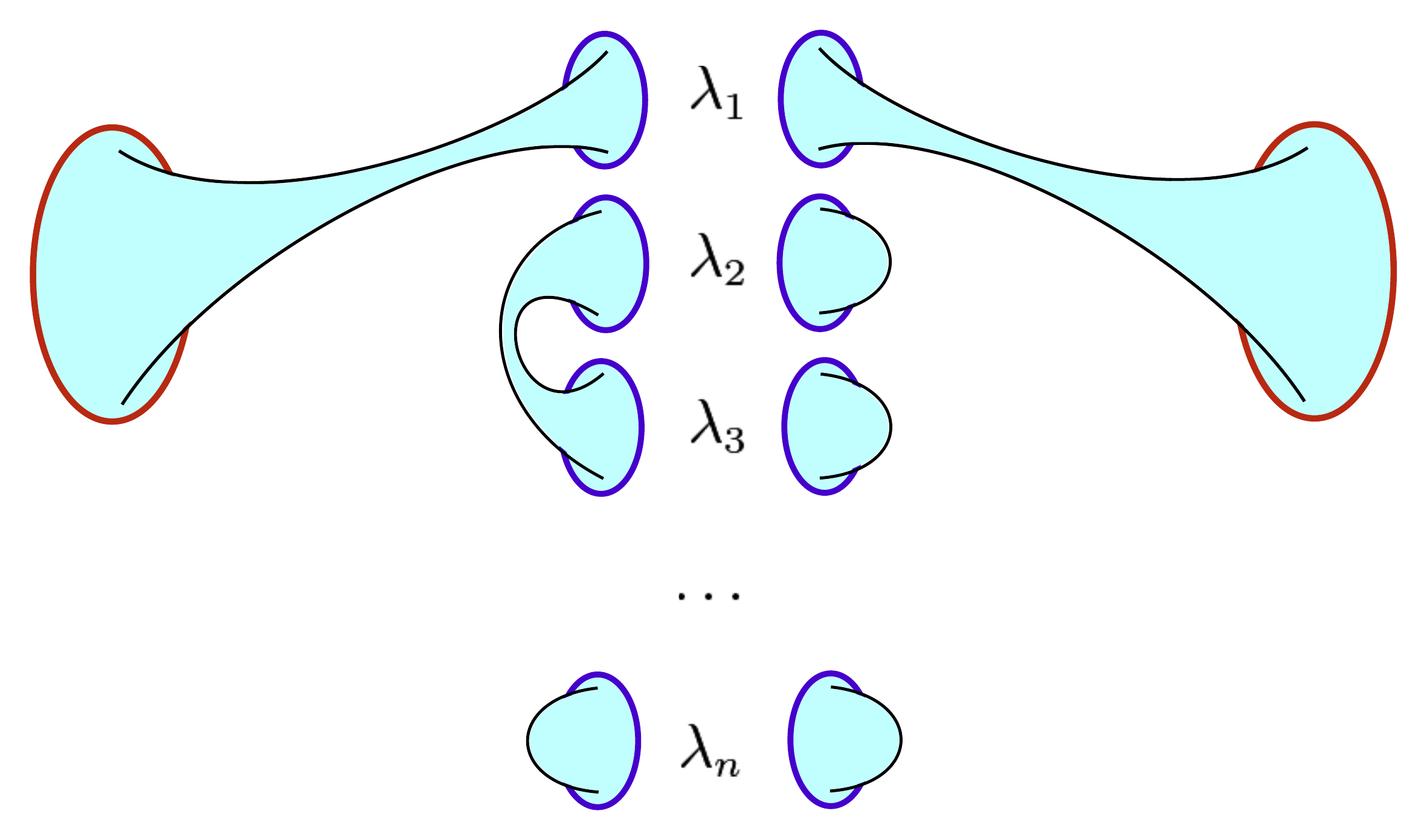}}
\end{equation}
This factorizing property is slightly surprising in the sense that the geometries on the right hand side are never counted in the original perturbative JT gravity path integral prescription for $\average{\rho(E_1)\rho(E_2)}_{\lambda_1\dots\lambda_n}$. One might have expected that connected contributions to the gravity analogue of a discrete system vanish. We find that instead they are non-zero, but their sum can be replaced by a sum over disconnected contributions.
\\
From the perspective of the matrix integral this disconnection is more intuitive. If we deplete a large energy region $\abs{\mathcal{I}} \gg 1$ then locally the randomness of the initial Hamiltonian is lost. Indeed, a close relation between geometric disconnection and lack of randomness is generally expected \cite{semiclassicalramp,sss2}.
\\
We note that \eqref{connect1} looks somewhat like introducing a ``complete set of baby universes" between $E_1$ and $E_2$ as hinted towards in \cite{phil}, though the status of these eigenbranes as ``states" in bulk JT gravity is at the moment unclear. More details regarding this section will be discussed in \cite{dissecting}.
\section{Concluding remarks}
\label{sect:concl}
It would be interesting to understand what these eigenbranes mean for a Lorentzian observer probing the gravitational bulk. Can he somehow obtain information about the branes hovering deep in the bulk? One way to work towards this would be to investigate boundary correlators in the matrix ensemble, see for example \cite{phil,Cotler:2019egt}. It would be valuable to understand if we can construct bulk observables within JT gravity as a sum over these more complicated geometries, using geodesic localizing, in analogy to the construction of local bulk observables in the disk version of JT gravity \cite{paper5,Mertens:2019bvy}. 
\\~\\
We end this work with three remarks.
\\~\\
\textbf{\emph{Boundary mergers}}
\\~\\
The Dirac delta's that appear in the exact answers for the spectral densities in section \ref{sect:32} have an a posteriori interpretation as eigenvalue boundaries merging with the asymptotic boundaries.\footnote{See for example \cite{1911.01659,ginspangmoore}.} Considering for example the last equality in \eqref{zz}, the first term can be read as counting Riemann surfaces which end on the merger of the two original boundaries of lengths $\beta_1$ and $\beta_2$, resulting in a boundary of total length $\beta_1+\beta_2$.
\\
Let us pretend here to take that interpretation seriously, and count Riemann surfaces that end on a merged boundary. It is convenient to introduce the JT gravity disk amplitude between a fixed length state $\ket{\beta}$ and a fixed energy state $\ket{E}$:\footnote{These are the states used in \cite{zhenbin,phil,harlowjafferis}, with $\ket{\beta}$ the Hartle-Hawking state of the JT gravity disk. We have:
\begin{equation}
    \ket{\beta}=\int_0^\infty d E\, e^{-\beta E}\, \rho_0(E)\ket{E},\quad \bra{E_1}\ket{E_2}=\frac{\delta(E_1-E_2)}{\rho_0(E)},\quad \bra{\beta_1}\ket{\beta_2}=Z_\text{JT}(\beta_1+\beta_2).
\end{equation}
}
\begin{equation}
    \bra{\beta}\ket{E}=\quad \raisebox{-8mm}{\includegraphics[width=25mm]{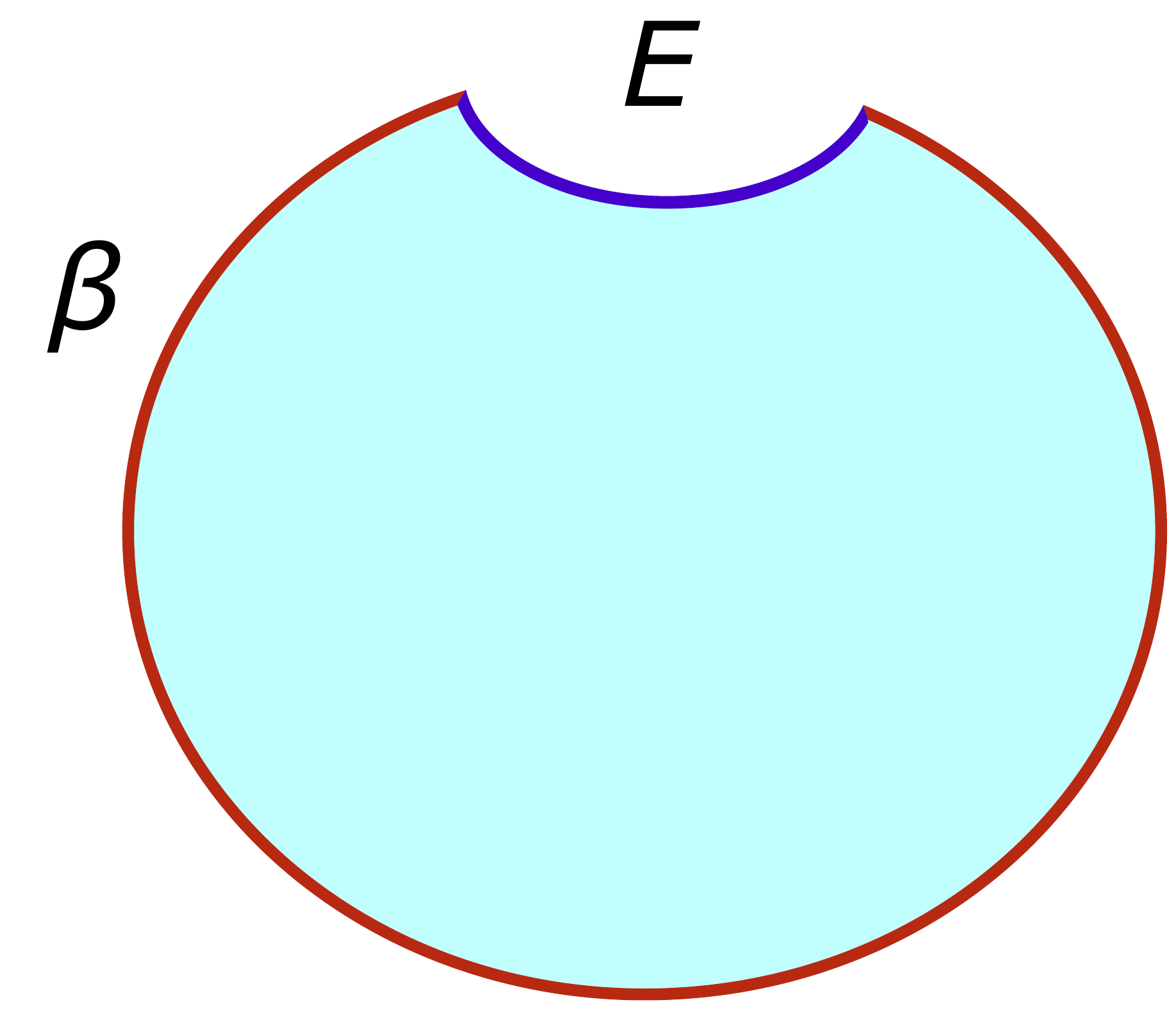}}\quad = e^{-\beta E}.
\end{equation}
The merger of an asymptotic disk with an eigenbrane results in the genus zero amplitude $\rho_0(\lambda)\bra{\beta}\ket{\lambda}$.\footnote{This is the inverse Laplace transform of the boundary with length $\beta_1+\beta_2$ with respect to $\beta_2$.}
This merged boundary can connect to the other eigenvalue boundaries, and develop handles. In taking the sum, the overlap $\bra{\beta}\ket{\lambda_i}$ is a spectator. We end up with a factor that cancels precisely the denominator in \eqref{perturbativerho}, and we are left only with $\bra{\beta}\ket{\lambda_i}$. As pointed out in section \ref{sect:fixing}, all other contributions to the JT gravity partition function add up to zero. This suggests the net gravitational effect of fixing all eigenvalues \eqref{spikes} in JT gravity is the following replacement:
\begin{equation}
    Z_\text{JT}(\beta)=\, \raisebox{-8mm}{\includegraphics[width=25mm]{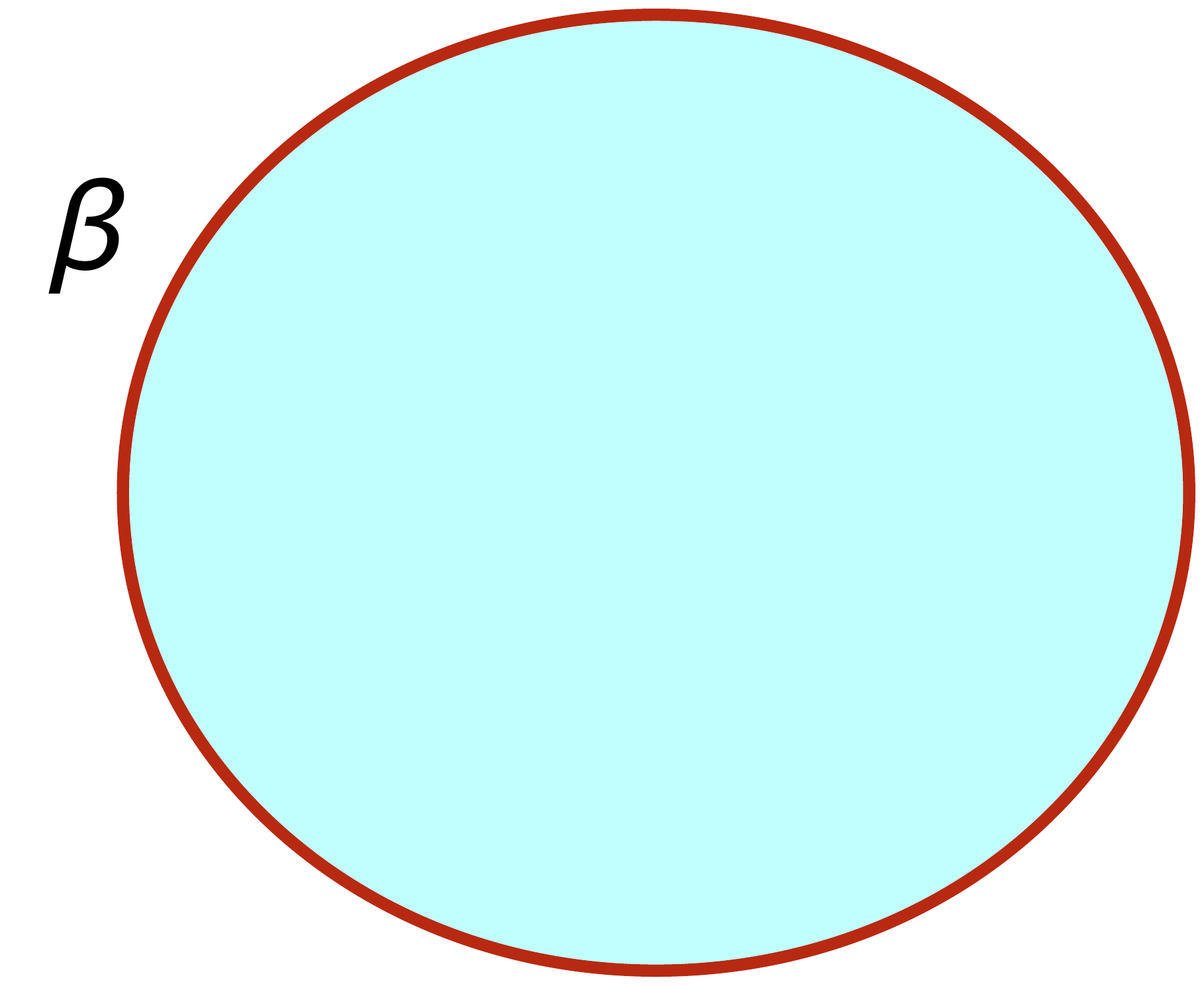}}\quad\to\quad \sum_{i=1}^N \raisebox{-8mm}{\includegraphics[width=25mm]{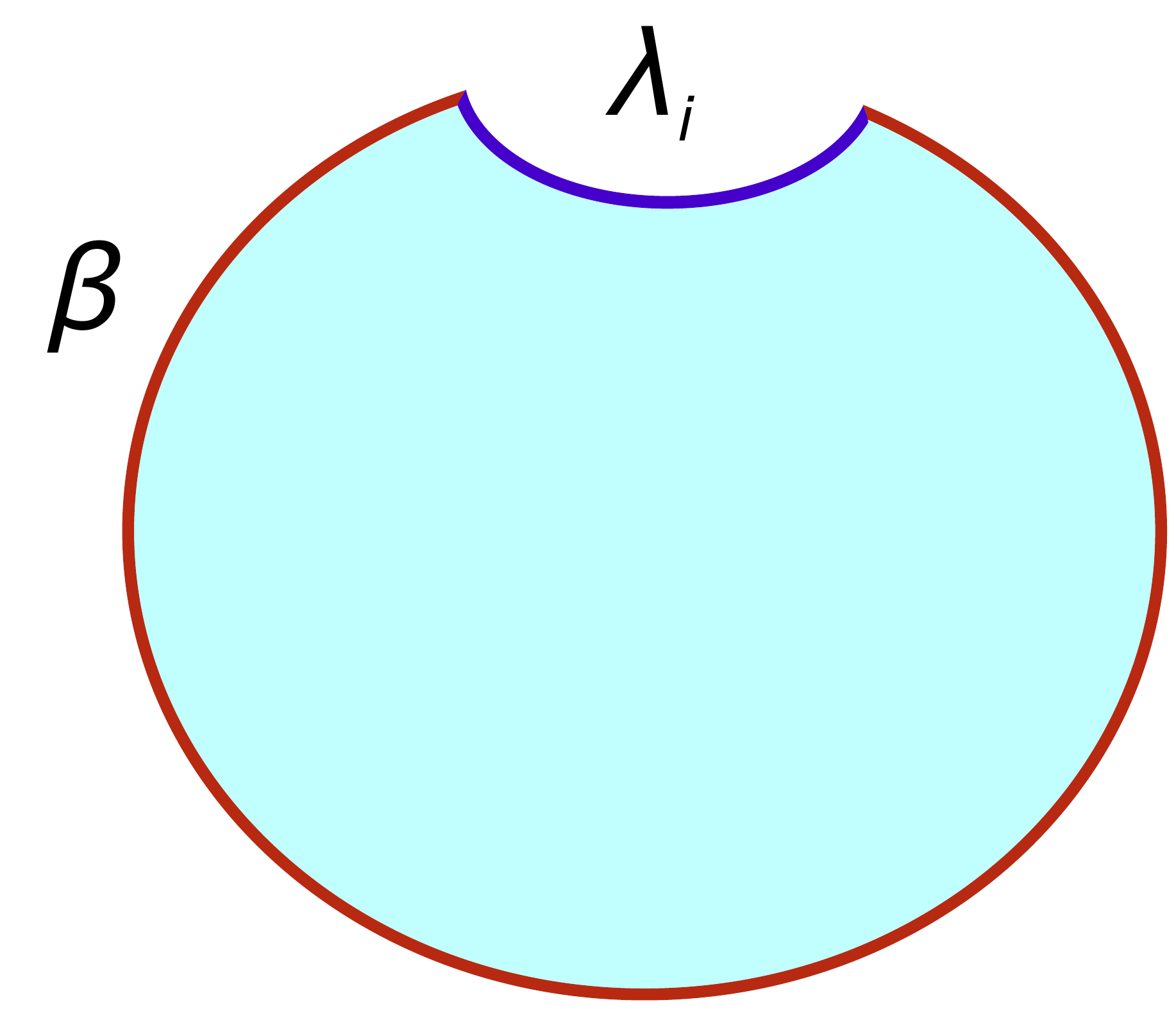}}\quad=\sum_{i=1}^N e^{-\beta\lambda_i}.\label{proposal0}
\end{equation}
It is tempting to imagine how this could extend to holographic correlation functions.\footnote{See \cite{phil,Cotler:2019egt} for recent discussions.} In JT gravity in its gauge-theoretic BF formulation, boundary correlators correspond to Wilson lines traversing the Riemann surfaces \cite{paper2,paper3,paper4,1905.02726,luca}. The Wilson line separates the Riemann surface into two disconnected pieces, each connected to a piece of boundary:\footnote{A similar such configuration with a vacuum Wilson line does not contribute to the JT gravity partition function, because the eigenvalues are chosen not to be degenerate, and merging two fixed energy boundaries to a fixed length boundaries results in an amplitude proportional to a Dirac delta on those energies.}
\begin{align}
    \raisebox{-9mm}{\includegraphics[width=30mm]{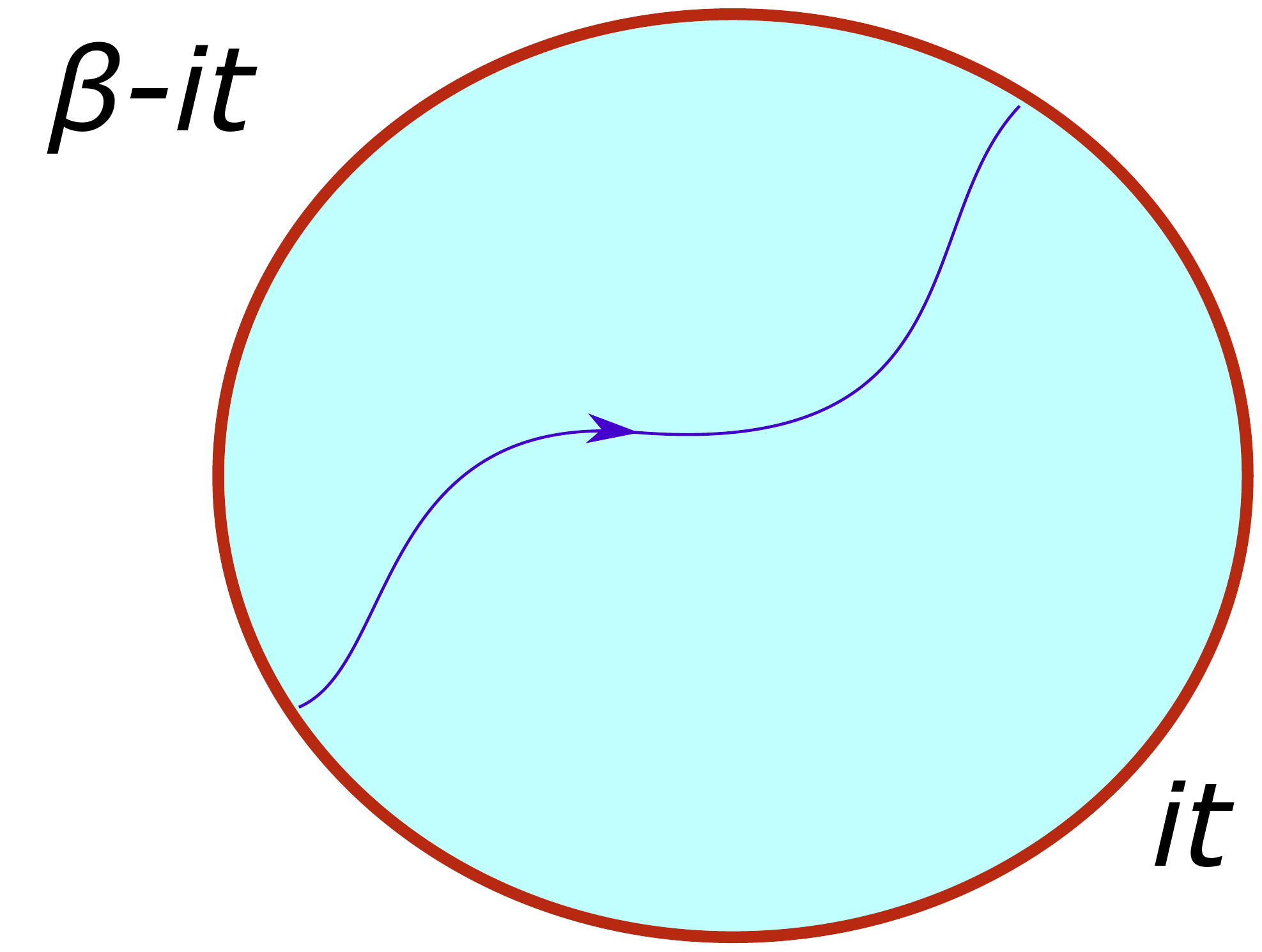}}\quad\to\quad &\sum_{i=1}^N\sum_{j=1}^N \raisebox{-11mm}{\includegraphics[width=30mm]{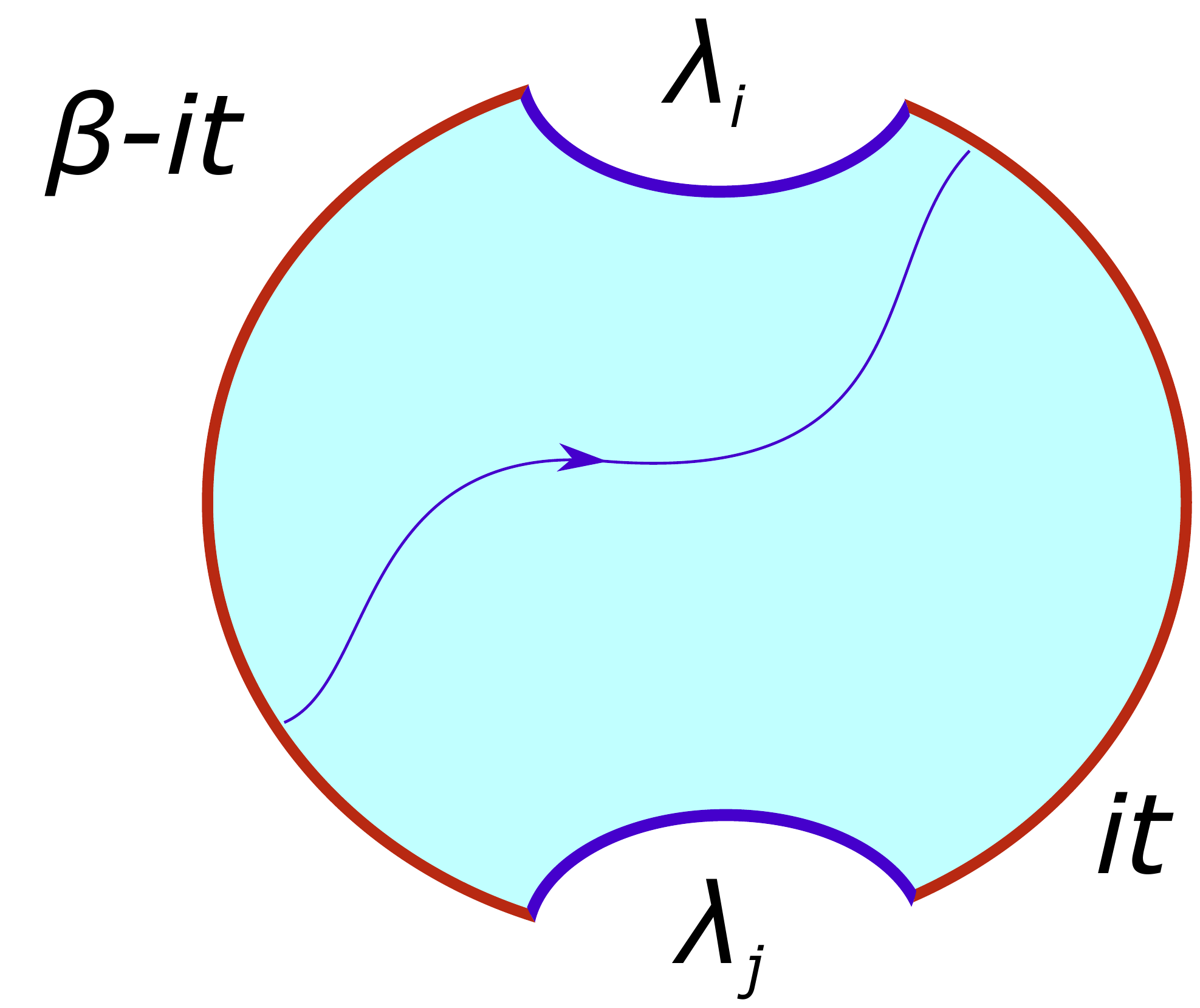}}\nonumber\\
    &=\sum_{i=1}^N\sum_{j=1}^Ne^{-(\beta-it)\lambda_i}\, e^{-it \lambda_j}\abs{\bra{\lambda_i}\mo\ket{\lambda_j}}^2.\label{proposal}
\end{align}
Even though the final expression \eqref{proposal} is the two point function of a discrete system for any operator $\mo$, it is nontrivial to find one that corresponds to a boundary-anchored Wilson line in the perturbative definition of JT gravity.
\\
One hint that \eqref{proposal} might be correct comes from the spectral form factor. This can be obtained from an analytic continuation of the Schwarzian boundary two-point function with $\beta-it\to \beta_1\gg 1$ and $it\to \beta_2\gg 1$. The Wilson line in \eqref{proposal} then effectively pinches off the disk.\footnote{The matrix elements $\bra{\lambda_i}\mo\ket{\lambda_j}$ are known in that context as the vertex functions \cite{Mertens:2017mtv} and go to a constant at low energy \cite{paper5}.} 
\\~\\
\textbf{\emph{Other ensembles}}
\\~\\
The analysis of this work is readily generalized to JT $\mathcal{N}=1$ supergravity \cite{stanfordwittenJT}. The general Altland-Zirnbauer ensembles are defined by the integration measure \cite{Altland:1997zz}:
\begin{equation}
d\mu(\lambda_1\hdots) = d\lambda_1\hdots d\lambda_n e^{-L \sum_{i=1}^{L}V(\lambda_i)} \prod_{i<j}(\lambda_j-\lambda_i)^\beta \lambda_i^{\frac{\alpha-1}{2}}, \qquad \lambda_i >0,
\end{equation}
in terms of two integers $(\alpha,\beta)$. It was shown in \cite{stanfordwittenJT} that these are related to JT supergravity either on orientable or orientable plus nonorientable surfaces depending on the choices of $\alpha$ and $\beta$. Taking $\beta=2$ and $\alpha=1$ is the orientable case (no time-reversal symmetry). For the nonorientable case (time-reversal symmetry), there is a possible divergence from the crosscap moduli space, with the exception of the case $\beta=2$ and $\alpha=0,2$. We focus hence only on these cases. 
\\
In analogy to \eqref{1levelspectral} and \eqref{2levelspectral} one has:
\begin{align}
	\nonumber \average{\rho(E)} &= E^{\frac{\alpha-1}{2}}\frac{1}{2\pi}\average{\psi^2(E)}_{\LL-1}, \\
    \average{\rho(E_1)\rho(E_2)} &= \frac{1}{(2\pi)^2}(E_1-E_2)^2E_1^{\frac{\alpha-1}{2}}E_2^{\frac{\alpha-1}{2}}\average{\psi^2(E_1)\psi^2(E_2)}_{\LL-2}\nonumber\\&\quad+\delta(E_1-E_2)E_1^{\frac{\alpha-1}{2}}\frac{1}{2\pi}\average{\psi^2(E_1)}_{\LL-1}.
\end{align}
These can be calculated efficiently by generalizing the brane computation of Appendix \ref{app:calculations} to including the crosscap $\text{Xcap}(e) = - \frac{\alpha-1}{4}\log(-E)$. Contributions to the spectral density then for example are:
\begin{equation}
     E^{\frac{\alpha-1}{2}}\average{\psi(e)\psi(e)}+\average{\psi(-e)\psi(-e)} = -\frac{1}{2E}\sin(-\frac{\pi \alpha}{2} + 2\pi\int_0^E d M \rho_0(M))=2\pi \rho_\text{nonp}(E), \nonumber
\end{equation}
and
\begin{equation}
 \nonumber E_2^{\frac{\alpha-1}{2}} \lim_{e_1\to e_2} \average{\psi(e_1)\psi(-e_2)}+\average{\psi(-e_1)\psi(e_2)} = \lim_{e_1\to e_2}\frac{\sinh(\text{Disk}(e_1)-\text{Disk}(e_2)))}{\sqrt{E}(e_1-e_2)}= 2\pi \rho_0(E).  
\end{equation}
In this case:
\begin{equation}
    \rho_0(E) = \frac{1}{2\pi}e^{S_0}\frac{\cosh 2\pi \sqrt{E}}{\sqrt{E}}.
\end{equation}
It is not difficult to considered fixed eigenvalues or conversely additional fixed energy boundaries in these models, one merely has to ship in the appropriate cluster functions for these $(\alpha,\beta)$ ensembles. Close to the spectral edge $\rho_0(E) \sim 1/\sqrt{E}$ JT supergravity reduces to the exactly solvable Bessel model, which is the super-analogue of the Airy model \cite{stanfordwittenJT}.
\\~\\
\textbf{\emph{A gravitational hint of ensemble averaging?}}
\\~\\
The statistical ensemble we found from the matrix integral, was interpreted gravitationally in terms of multiple boundaries. Here we illustrate that starting with gravity directly, one can get some hints of this underlying ensemble.
\\
We start from a property of the $n$-boundary correlator in JT gravity \eqref{ncorrdef}:
\begin{equation}
    Z_\text{JT}(\beta_1\dots \beta_n)\supset e^{\chi(g,n)S_0}\int_0^\infty b_1 d b_1\, Z_\text{JT}(\beta_1,b_1)\dots \int_0^\infty b_n d b_n\, Z_\text{JT}(\beta_2,b_n)\, V_{g,n}(b_1\dots b_n).
\end{equation}
Let us take the length of one of the boundaries to zero. The single-trumpet partition function $Z_\text{JT}(\beta)$ \eqref{sch} becomes a delta-function for $\beta\to 0$. Taking $\beta_1\to 0$ therefore localizes on spacetimes where the neck length $b_1$ vanishes. Due to the twist factor $b_1$ and the polynomial behavior of the Weil-Petersson volumes, we see that every perturbative contribution vanishes except for the case when the $\beta_1$-boundary is capped off by a disk. We end up with:\footnote{This limit should be taken with care in full JT gravity, since the asymptotic boundary length $L \equiv \beta/\epsilon \to +\infty$ in JT gravity \cite{1606.01857,1606.03438}, where afterwards we let $\beta\to 0$.}
\begin{equation}
    Z_\text{JT}(0,\beta_1\dots \beta_n)=Z_\text{JT}(0)Z_\text{JT}(\beta_1\dots \beta_n).\label{pinch}
\end{equation}
Doing an $n$-fold inverse Laplace transform of this equation, we find:
\begin{equation}
    \int_0^\infty d\lambda\,\rho(\lambda,\lambda_1\dots \lambda_n)=\rho(\lambda_1\dots \lambda_n)\int_0^\infty d\lambda\rho(\lambda)=\rho(\lambda_1\dots \lambda_n) Z_\text{JT}(0).
\end{equation}
Recursively, one gets from this:
\begin{align}
    \frac{1}{Z_\text{JT}(0)^n}\int_0^\infty d\lambda_1\dots d\lambda_n\, \rho(E,\lambda_1\dots \lambda_n)&= \rho(E), \label{rhoe} \\
    \frac{1}{Z_\text{JT}(0)^n}\int_0^\infty d\lambda_1\dots d\lambda_n\,\rho(\lambda_1\dots \lambda_n)&= 1.\label{sphere}
\end{align}
This suggests to think of $\rho(\lambda_1\dots \lambda_n)\,Z_\text{JT}(0)^{-n} = w(\lambda_1\dots\lambda_n)$ as the weight function of a statistical ensemble. This is strengthened by \eqref{rhoe} and its generalization to multiple $E_i$: correlators in JT gravity can be calculated as averages in this statistical ensemble. In particular the observable that calculates $\rho(E)$ is extracted from \eqref{rhoe} as:
\begin{equation}
    \int_\mathcal{C} d\lambda_1\dots d\lambda_n\, w(\lambda_1\dots\lambda_n)\, \rho(E)_{\lambda_1\dots \lambda_n}=\rho(E), \qquad \label{forget}
\rho(E)_{\lambda_1\dots\lambda_n}=\frac{\rho(E,\lambda_1\dots \lambda_n)}{\rho(\lambda_1\dots \lambda_n)}.
\end{equation}
This corresponds to the quantity we considered in the main text.
\section*{Acknowledgements}
We thank Phil Saad for a useful discussion. AB and TM gratefully acknowledge financial support from FWO Vlaanderen. 
\appendix

\section{Some brane calculations}\label{app:calculations}
In this appendix, we calculate brane pair correlators of the type $\average{\psi^2(E_1)\dots \psi^2(E_n)}$ in JT gravity, with a single brane defined as \eqref{branedef}. As discussed in the main text \eqref{rbrane}, this is an efficient way to calculate objects such as $R(E)$ and $R(E_1,E_2)$.
\\~\\
We can rewrite the brane operator \eqref{branedef} as:
\begin{equation}
    \psi(E)=e^{-\frac{LV(E)}{2}}\prod_{i=1}^L(E-\lambda_i)=\exp(-\frac{L V(E)}{2}+\Tr \log (E-M) ).\label{exp}
\end{equation}
The operator in the exponential corresponds to the insertion of an unmarked fixed energy boundary in JT gravity \cite{sss2}:
\begin{equation}
    \text{Disk}(E)=-\frac{L V(E)}{2}+\Tr \log (E-M)=-\int_\mathcal{C} \frac{d\beta}{\beta}\, e^{\beta E}\, Z(\beta).\label{fzzt}
\end{equation}
This is the precise analogue of an unmarked FZZT boundary brane in Liouville theory \cite{0001012,teschner}. Equation \eqref{exp} is slightly misleading in combination with \eqref{fzzt} though. The original brane correlator \eqref{branedef} is an analytic function of $E$, whereas the FZZT brane \eqref{fzzt} has a discontinuity on the positive real axis. Consequently, to each energy $E$ there correspond two different FZZT boundaries in gravity, depending on how we approach the real axis. This is equivalent to specifying the value $\sqrt{-E}$ for $E>0$. Let us introduce a new variable $e=i\sqrt{E}$, then $\sqrt{-E}=\pm e$ for $E>0$. Depending on this sign, exponentiating the FZZT boundary \eqref{fzzt} gives two distinct gravitational brane correlators $\average{\psi(\pm e)}$.
\\
This raises the question which gravitational brane corresponds to inserting the brane operator \eqref{branedef} in the matrix integral. The answer was given in \cite{0408039}.\footnote{See also \cite{sss2}.} The brane correlators have an exact expression for finite $e^{S_0}$ as a Kontsevich matrix integral, or an appropriate JT gravity generalization thereof.\footnote{See \cite{1911.01659}.} For $e^{S_0}\gg 1$ we can use the method of Laplace on this Kontsevich matrix integral. Depending on whether the energy parameters $E_1\dots E_n$ are positive or negative, different saddles contribute due to Stokes phenomena. Each such saddle, and the loop corrections around it, correspond to a gravitational brane. It turns out that for all energies $E_1\dots E_n$ positive, we need to sum over all possible corresponding gravitational branes with equal weight. For each such saddle, if we furthermore take $E\gg e^{-2S_0/3}$ only the exponentiation of disks and annuli contributes significantly.\footnote{Higher genus surfaces give multiplicative contributions of the type $e^{\chi e^{S_0}\dots}$, with $\chi<0$ and the $\dots$ polynomials in $1/E_1$ multiplied with $(E_1\dots)^{-1/2}$ following a modification of the calculations in section \ref{sect:31} to unmarked boundaries.}
\\
We will be working throughout in the regime $e^{S_0}\gg 1$. In sections \ref{onepairr} and \ref{twopairr} we furthermore assume $E\gg e^{-2S_0/3}$. In section \ref{app:airy} we calculate the correlators close to the spectral edge using the Airy model. The Airy calculations are exact for any $e^{S_0}$ and by construction coincide with the JT gravity answers for $E\ll 1$. For $e^{S_0}\gg 1$ these regions overlap, and we have an exact answer for all $E$.
\subsection{One pair}\label{onepairr}
Consider now the calculation of $R(E)$, corresponding to a single brane pair \eqref{1levelspectral}. Summing all saddles results in:
\begin{equation}
    \average{\psi^2(E)}=\average{\psi(e)\psi(e)}+\average{\psi(e)\psi(-e)}+\average{\psi(-e)\psi(e)}+\average{\psi(-e)\psi(-e)}.\label{branepair}
\end{equation}
As explained above and in \cite{0408039,sss2} only disks and annuli are significant in the regime we are focusing on:
\begin{align}
\label{exponential}
  \average{\psi(e_1)\psi(e_2)} 
		\approx e^{\text{Disk}(e_1)+\text{Disk}(e_2)+\frac{1}{2}\text{Ann}(e_1,e_1)+\frac{1}{2}\text{Ann}(e_2,e_2)+\text{Ann}(e_1,e_2)}.
\end{align}
One obtains the fixed energy disk-and annuli as Laplace transforms of the leading JT gravity fixed length disk \eqref{diskspec} and annuli amplitude \eqref{annuluscalc}, as in \eqref{fzzt}. The result is \cite{sss2}:
\begin{equation}
    \text{Disk}(\pm e)=\pm i \pi \int_0^E d M \rho_0(M),\quad \text{Ann}(e_1,e_2)=-\ln(e_1+e_2).\label{mdisk}
\end{equation}
Note that these indeed both depend explicitly on the sign $\pm e$. We get:
\begin{equation}
    \average{\psi(e_1)\psi(e_2)}=\frac{1}{2\sqrt{e_1}\sqrt{e_2}(e_1+e_2)}e^{\text{Disk}(e_1)+\text{Disk}(e_2)}.\label{pole}
\end{equation}
Using
\begin{align}
    \average{\psi(e)\psi(e)}+\average{\psi(-e)\psi(-e)} &= -\frac{1}{2E}\cos(2\pi\int_0^E d M \rho_0(M))=2\pi \rho_\text{nonp}(E), \nonumber \\
\lim_{e_1\to e_2} \average{\psi(e_1)\psi(-e_2)}+\average{\psi(-e_1)\psi(e_2)} &= \lim_{e_1\to e_2}\frac{\sinh(\text{Disk}(e_1)-\text{Disk}(e_2)))}{\sqrt{E}(e_1-e_2)}=\frac{1}{\sqrt{E}}\partial_e\text{Disk}(e),\end{align}
we end up with:
\begin{equation}
    \average{\psi^2(E)}=2\pi \rho_0(E)-\frac{1}{2E}\cos(2\pi\int_0^E d M\rho_0(M))=2\pi R(E).
\end{equation}
This matches the result of the resolvent-based brane dipole calculation of $R(E)$ in \cite{sss2}.

\subsection{Two pair}\label{twopairr}
Next we calculate the two-brane-pair correlator $\average{\psi^2(E_1)\psi^2(E_2)}$. For notational purposes consider $\average{\psi^2(E)\psi^2(K)}$ with $e=i\sqrt{E}$ and $k=i\sqrt{K}$. Summing the 16 saddles gives:
\begin{equation}
    \average{\psi^2(E)\psi^2(K)} = \sum_{\text{signs}}\average{\psi(\pm e)\psi(\pm e)\psi(\pm k)\psi(\pm k)}.\label{twopair}
\end{equation}
The generic brane correlator is similar to \eqref{exponential}:
\begin{equation}
    \average{\prod_{i} \psi(e_n)}
		= \prod_{i} e^{\text{Disk}(e_i)+\frac{1}{2}\text{Ann}(e_i,e_j)}\prod_{j\neq i} e^{\text{Ann}(e_i,e_j)}.
\end{equation}
Using \eqref{mdisk} we obtain:
\begin{align}
    \nonumber\average{\psi(e_1)\psi(e_2)\psi(e_3)\psi(e_4)}&\approx \frac{1}{4\sqrt{e_1}\sqrt{e_2}\sqrt{e_3}\sqrt{e_4}}\\
    &\times \frac{e^{\text{Disk}(e_1)+\text{Disk}(e_2)+\text{Disk}(e_3)+\text{Disk}(e_4)}}{(e_1+e_2)(e_1+e_3)(e_1+e_4)(e_2+e_3)(e_2+e_4)(e_3+e_4)}.\label{total}
\end{align}
It is now a straightforward but somewhat tedious task to evaluate \eqref{twopair}. The 16 terms fall in three classes. Firstly, there are 4 terms where the signs match within each brane pair:
\begin{align}
\label{co1}
  \nonumber \sum_{s_a,s_b = \pm} \average{\psi(s_a e)\psi(s_a e)\psi(s_b k)\psi(s_b k)}&=\\ \frac{2\cosh(2\disk(e)\hspace{-1mm}-\hspace{-1mm}2\disk(k))}{16 E K(e-k)^4}&+\frac{2\cosh(2\disk(e)\hspace{-1mm}+\hspace{-1mm}2\disk(k))}{16 E K(e+k)^4}.
\end{align} 
Secondly, there are 8 mixed terms:
\begin{align}
   &\sum_{s_a,s_b = \pm} \average{\psi(s_a e)\psi(s_a e)\psi(s_b k)\psi(-s_b k)} + \sum_{s_a,s_b = \pm} \average{\psi(s_a e)\psi(-s_a e)\psi(s_b k)\psi(s_b k)} \nonumber \\
	\nonumber &= \frac{4\pi^2}{(E-K)^2}\rho_0(K)\,\rho_\text{nonp}(E) + \frac{4\pi^2}{(E-K)^2}\rho_0(E)\,\rho_\text{nonp}(K) + \frac{\sinh(2\disk(k))-\sinh(2\disk(e))}{i\sqrt{-e^2}\sqrt{-k^2}(e^2-k^2)^3}.
\end{align}
The remaining 4 terms have opposite signs within each brane pair. This terms requires a double use of L'H\^opital's rule:
\begin{align}
   \nonumber \sum_{s_a,s_b = \pm} \average{\psi(s_a e)\psi(-s_a e)\psi(s_b k)\psi(-s_b k)}= \frac{4\pi^2}{(E-K)^2}\rho_0(E)\rho_0(K) - \frac{(K+E)}{(K-E)^4}\frac{1}{\sqrt{E}\sqrt{K}}.
\end{align} 
One recognizes the first term as the product of two perturbative disks and the second term as the perturbative annulus \eqref{annulus}. \\
Adding these three contributions and multiplying by $(E-K)^2/4\pi^2$ gives the exact pair density correlator for $e^{S_0}\gg 1$ away from the spectral edge. We can distill from the exact answer $T(E,K)$ in JT gravity:
\begin{align}
    R(E,K) =& R(E)R(K) - \frac{(K+E)}{4\pi^2 (K-E)^2}\frac{1}{\sqrt{E}\sqrt{K}} + \frac{\sinh(2\disk(k))-\sinh(2\disk(e))}{4\pi^2 i\sqrt{E}\sqrt{K}(K-E)} \nonumber \\ 
	&+ \frac{2\cosh(2\disk(e))\cosh(2\disk(k))}{4\pi^2(K-E)^2}\nonumber\\
	&- \frac{\sinh(2\disk(e))\sinh(2\disk(k))(K+E)}{4\pi^2(K-E)^2}\frac{1}{\sqrt{E}\sqrt{K}}. \label{clusteraxact}
\end{align}
This connected contribution contains the perturbative annulus as first term. The remainder is the non-perturbative contribution to the two-holed sphere (i.e. the annulus).
\\
To uncover the GUE structure \eqref{cluster}, we focus on $\abs{E-K}\ll 1$.\footnote{We introduce $E_-=(E-K)/2$ and $E_+=(K+E)/2$.} This simplifies things considerably:
\begin{align}
    \frac{\sinh(2\disk(k))-\sinh(2\disk(e))}{i\sqrt{E K}(K-E)^3}&=-\frac{8\pi^2}{(E-K)^2} \rho_0(E_+)\,\rho_\text{nonp}(E_+)+\mo\left(\frac{1}{E-K}\right).
\end{align}
Furthermore:
\begin{align}
    \nonumber&\frac{1}{16 E K}\frac{2\cosh(2\disk(e)-2\disk(k))}{(e-k)^4}-\frac{(K+E)}{(K-E)^4}\frac{1}{\sqrt{K E}}\\\nonumber &=-\frac{4}{(E-K)^4} \sin^2(\pi \int_K^E d M\, \rho_0(M))-\frac{2}{E_+^2 (E-K)^2}\sin^2(\pi \int_K^E d M\, \rho_0(M))+\mo\left(\frac{1}{E-K}\right).
\end{align}
Collecting everything, we find for $\abs{E_1-E_2}\ll 1$:
\begin{align}
    \nonumber R(E_1,E_2)&=\rho_0(E_1)\rho_0(E_2)+\rho_0(E_1)\rho_\text{nonp}(E_2)+\rho_0(E_2)\rho_\text{nonp}(E_1)-2\rho_0(E_+)\rho_\text{nonp}(E_+)\\
    \nonumber &-\rho_0(E_1)\rho_0(E_2)\sinc^2 \pi \rho_0(E_+) (E_1-E_2)-\frac{1}{2E_+^2}\sin^2 \pi \rho_0(E_+)(E_1-E_2)\\
    &=\rho_0(E_1)\rho_0(E_2)(1-\sinc^2 \pi \rho_0(E_+) (E_1-E_2))+R_\text{wiggle}(E_1,E_2).
\end{align}
The first term is the well known GUE result. the second term is small and oscillatory. It is the analogue of the wiggles \eqref{wiggle} in $R(E)$. For the purpose of our story in the main text, these wiggles are negligible but for the fact that $R_\text{wiggle}(E_1,E_1)=0$, such that $R(E_1,E_1)=0$ as demanded by eigenvalue repulsion in the ensemble. 
\subsection{More pairs}
This procedure readily extends to generic $n$.
\\
The perturbative contribution is found by picking opposite signs within each brane pair, only then is there no oscillatory contribution. For example for $n=3$ after a tedious three-fold application of L'H\^opital's rule one recognizes the perturbative disks and annuli:
\begin{align}
&R(E,K,M)_\text{pert}\nonumber \\
=&\, \frac{(e^2-k^2)^2(k^2-m^2)^2(m^2-e^2)^2}{8\pi^3} \sum_{\text{signs}}\average{\psi(\pm e)\psi(\mp e)\psi(\pm k)\psi(\mp k)\psi(\pm m)\psi(\mp m)} \nonumber \\ 
=&\, \rho_0(E)\rho_0(K)\rho_0(M) + \rho_0(E) \frac{(k^2+m^2)}{(k^2-m^2)^2}\frac{1}{\sqrt{-k^2}\sqrt{-m^2}}\nonumber\\ 
&+ \rho_0(K) \frac{(m^2+e^2)}{(m^2-e^2)^2}\frac{1}{\sqrt{-m^2}\sqrt{-e^2}}+\rho_0(M) \frac{(e^2+k^2)}{(e^2-k^2)^2}\frac{1}{\sqrt{-e^2}\sqrt{-k^2}} \nonumber\\
=&\nonumber \quad\raisebox{-13mm}{\includegraphics[width=30mm]{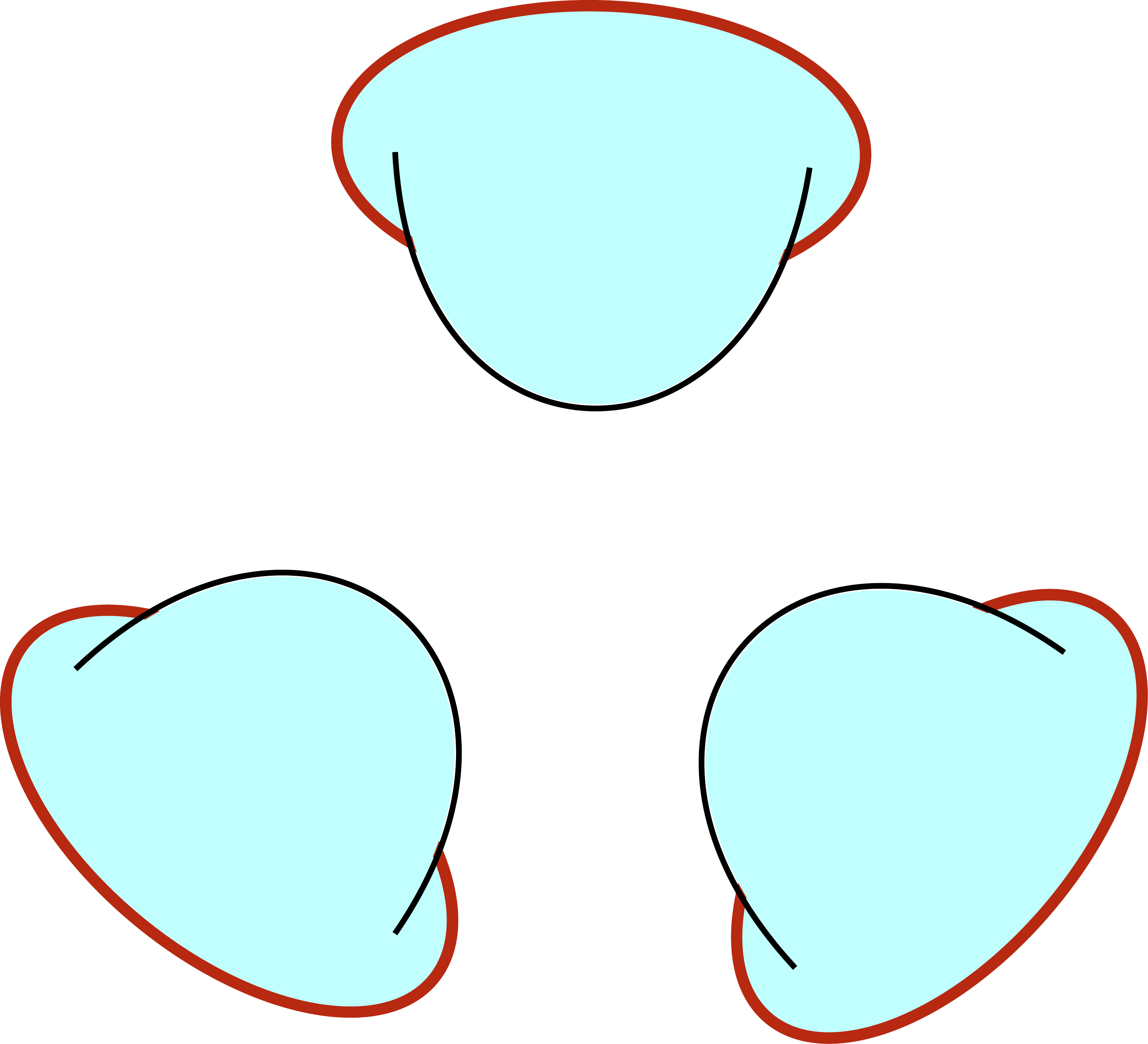}}\quad+\quad \raisebox{-13mm}{\includegraphics[width=30mm]{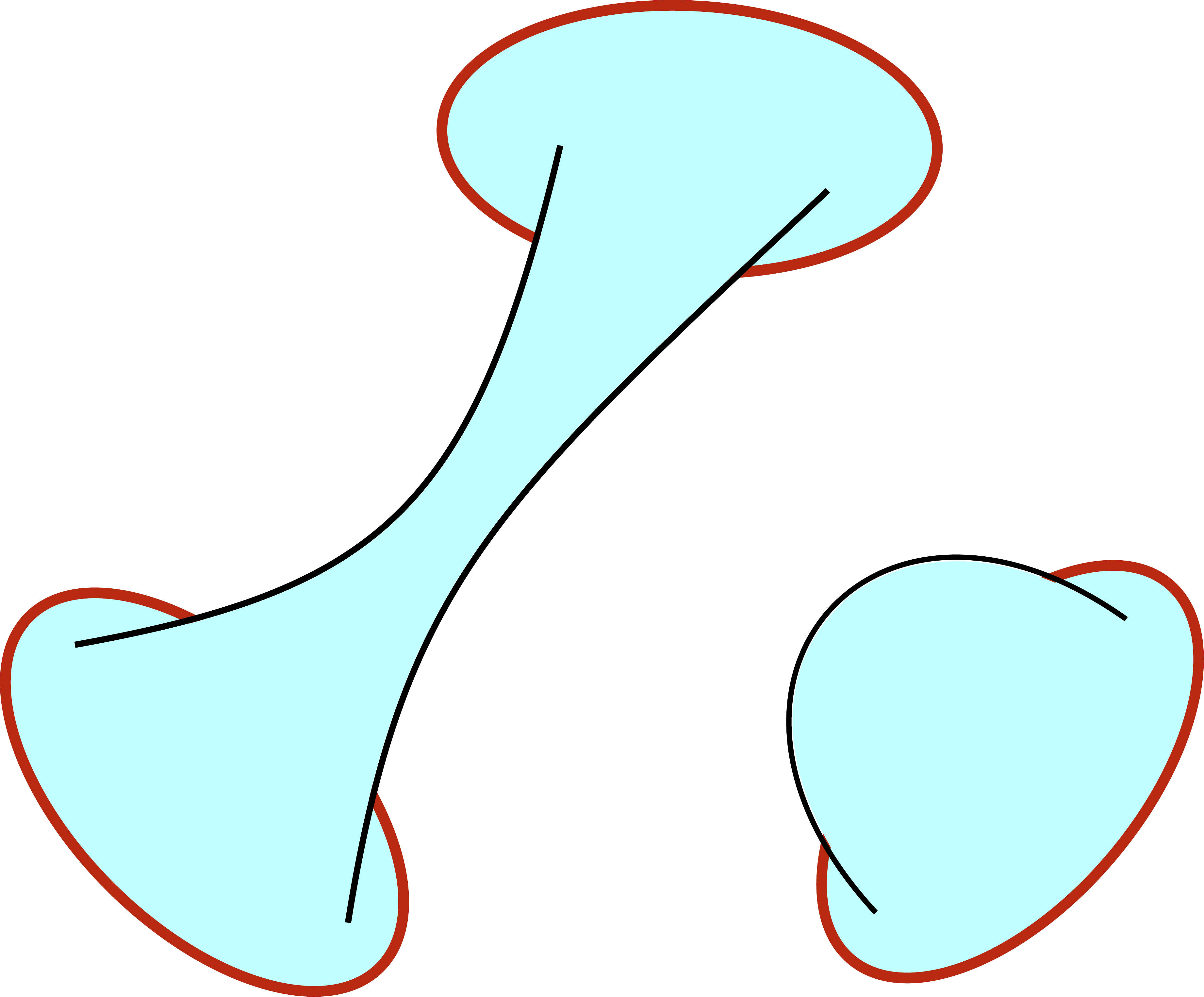}}\quad +\quad\raisebox{-13mm}{\includegraphics[width=30mm]{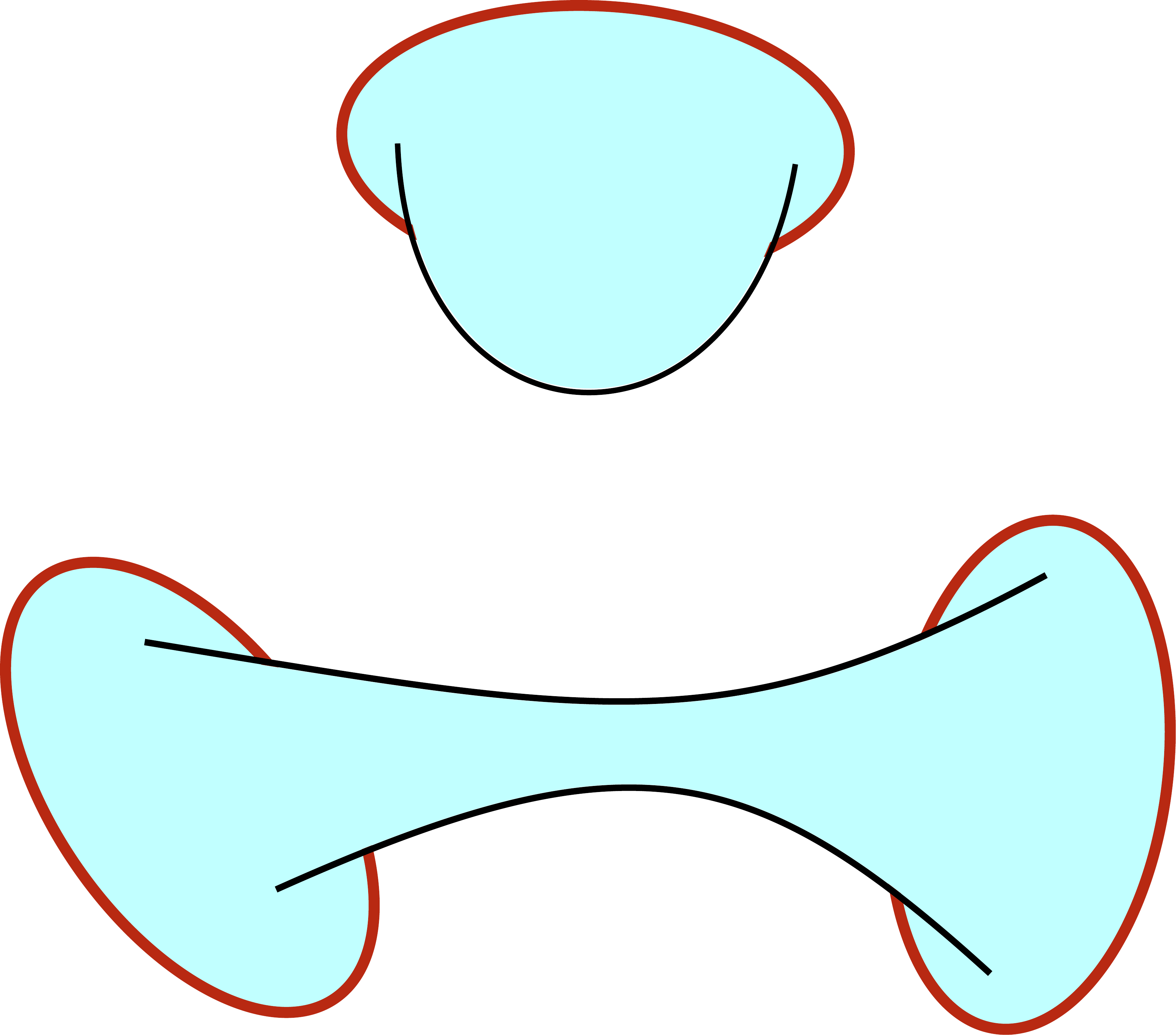}}\quad +\quad\raisebox{-14mm}{\includegraphics[width=30mm]{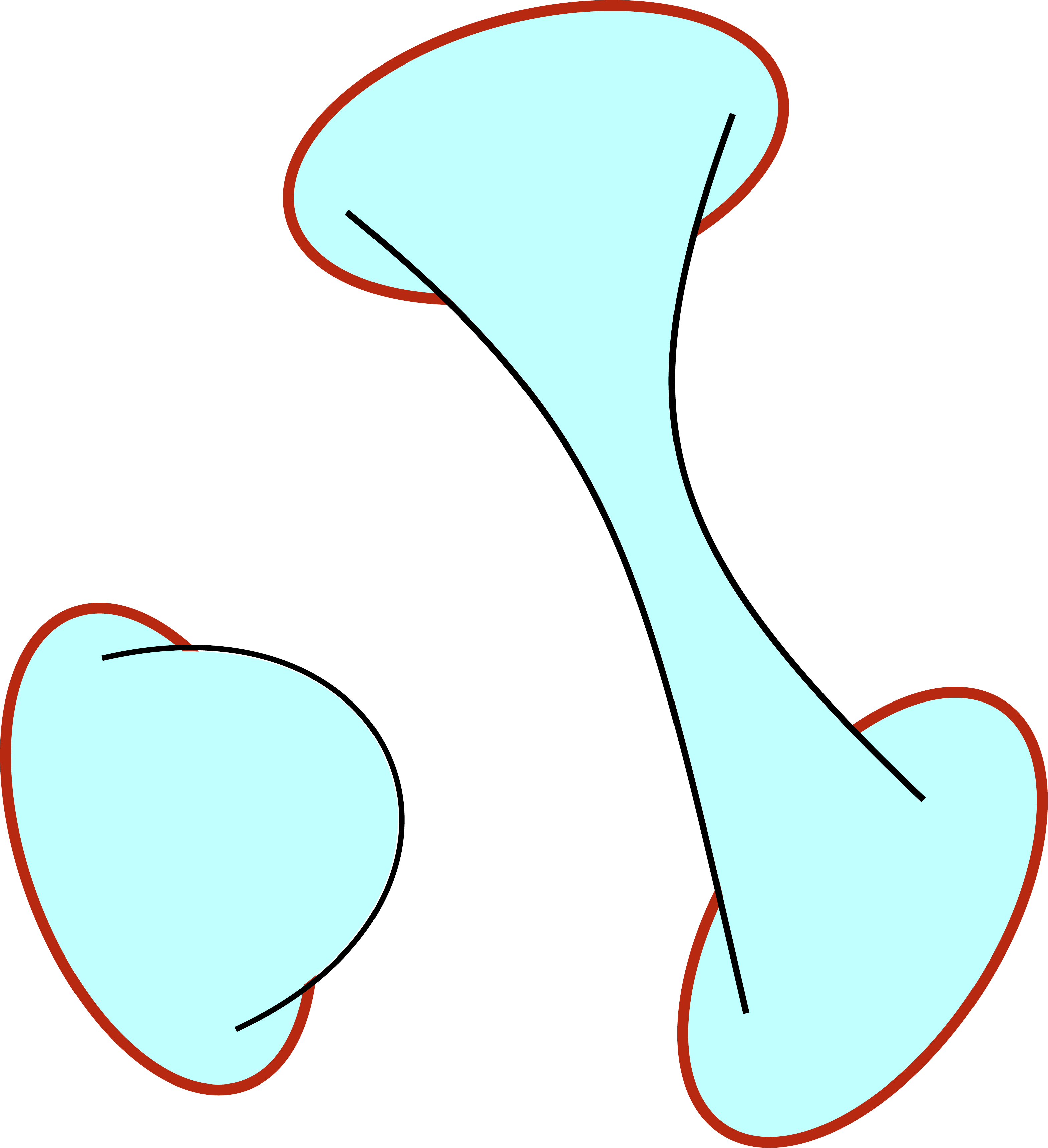}}
\end{align}
It is reassuring to see these and only these perturbative contributions appear. Notice for example that there is no perturbative three holed sphere contribution, nor are there handle-body geometries. This is consistent, as discussed in section \ref{sect:31} those geometries don't contribute significantly. On the other hand, the full $R(E,K,M)$ does for example contain the nonperturbative corrections associated to the genus expansion seeded by the three-holed sphere, which \emph{are} significant.\footnote{We could find more perturbative contributions by including the exponentials of these surfaces in the brane correlators such as \eqref{exponential}.}
\subsection{Fixing eigenvalues near the spectral edge}\label{app:airy}
Close to the spectral edge $\abs{E}\ll 1$, JT gravity reduces to topological gravity or the Airy model with spectral density:\footnote{We have rescaled the energy, removing the $e^{S_0}$ dependence here.}
\begin{equation}
    \rho_0(E)=\frac{\sqrt{E}}{\pi}.
\end{equation}
This theory is identical to the $(2,1)$ minimal string. The $(p,1)$ minimal strings are topological, and for these models the multi-brane correlators can be calculated exactly for any value of the string coupling.\footnote{It is solvable because we can solve the two coupled differential equations that define the single brane correlator. This function is known as a Baker-Akhiezer function of the KP hierarchy. For more on that see for example \cite{0408039} or the lecture notes \cite{dijkgraaf}.} This is the content of formula (1.11) in \cite{0408039}. In the case of the $(2,1)$ minimal string, we have:
\begin{equation}
    \average{\psi(x)}=\airy(x), \qquad x=-E. \label{airy}
\end{equation}
Multi-brane correlators are then calculated as formula (1.11) in \cite{0408039}:\footnote{We have $\average{\psi(x,\tau)}=\airy(x+\tau)$, therefore $\partial_\tau=\partial_x$, which is one of the two differential equations that defines the Baker-Akhiezer function. The other one is the Airy equation. By rescaling the energy axis we can eliminate the $\tau$-dependence.}
\begin{equation}
    \average{\prod_{i=1}^n\psi(x_i)}=\frac{\Delta^{1/2}(\partial_1\dots \partial_n)}{\Delta^{1/2}(x_1\dots x_n)}\prod_{i=1}^n\average{\psi(x_i)}.\label{main}
\end{equation}
It is again straightforward, but slightly tedious to calculate the multi-brane-pair correlators that get the Airy cluster functions. We we'll show how this goes for $R(E)$ and $R(E_1,E_2)$, and investigate the Airy spectral density with one fixed eigenvalue $\average{\rho(E)}_\lambda$.
\\~\\
For the two-brane correlator, we have:
\begin{equation}
    \average{\psi(x_1)\psi(x_2)}=\frac{\average{\psi(x_1)}'\average{\psi(x_2)}-\average{\psi(x_1)}\average{\psi(x_2)}'}{x_1-x_2}.
\end{equation}
Setting $x_1\to x_2$, one finds:
\begin{equation}
    \average{\psi(x)^2}=\average{\psi(x)}''\average{\psi(x)}-\average{\psi(x)}'^2.
\end{equation}
Inserting the solution \eqref{airy}, and using the Airy equation $\airy''(x)=x\airy(x)$, this becomes:
\begin{equation}
    \average{\psi(x)^2}=x\airy(x)^2-e^{2S_0}\airy'(x)^2.
\end{equation}
This is proportional to the Airy spectral density:\footnote{The normalization of the wavefunction \eqref{airy} is chosen different from that in \eqref{pole}, hence the different proportionality factor.}
\begin{equation}
    R(E)=-\average{\psi^2(-E)} = \quad\raisebox{-15mm}{\includegraphics[width=50mm]{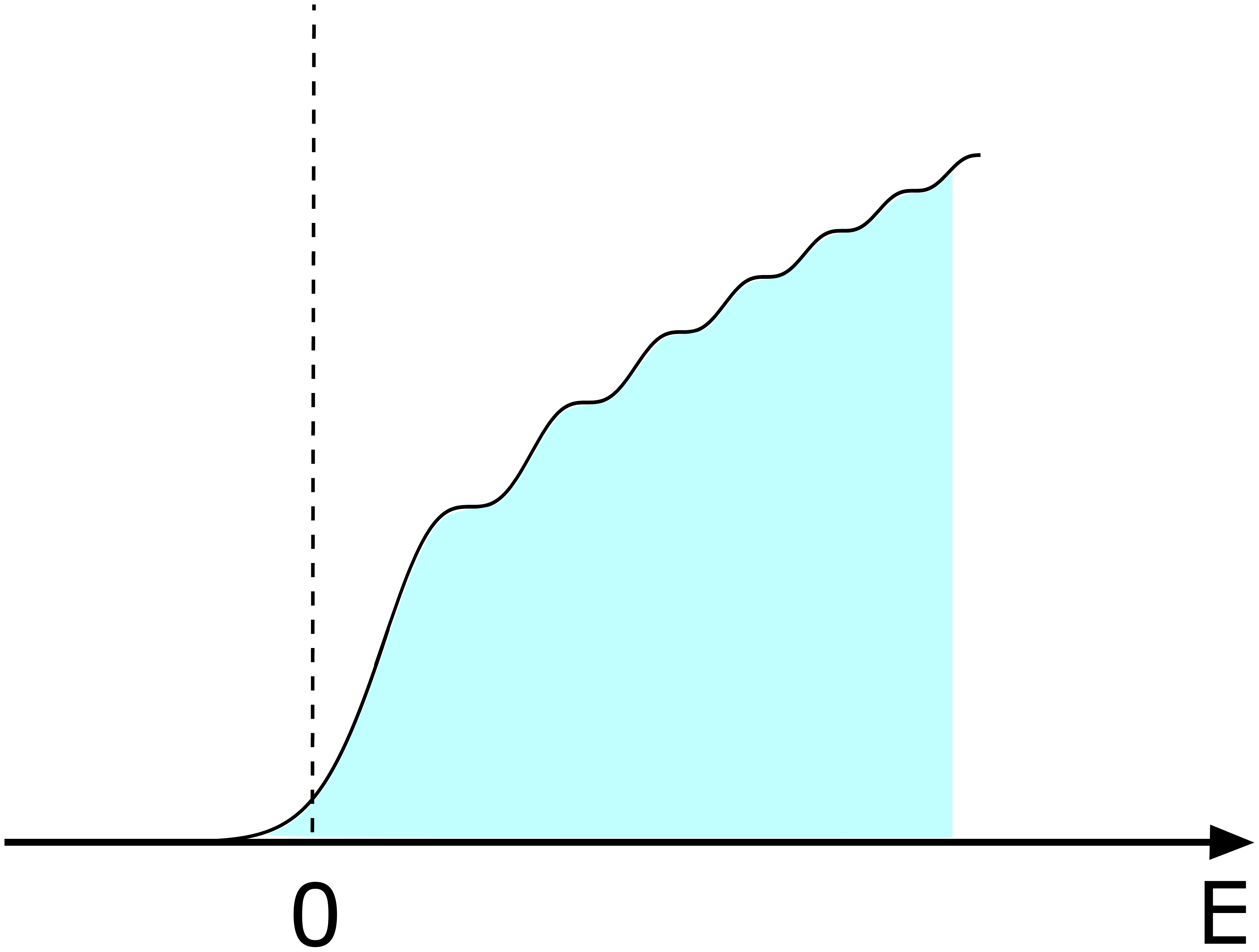}}\label{rairy} 
\end{equation}
To calculate the two-brane-pair correlator, we are led to consider \eqref{main}:
\begin{align}
    \nonumber &\average{\psi(x_1)\psi(x_2)\psi(x_3)\psi(x_4)}\\
    \nonumber&=\frac{(\partial_1-\partial_2)(\partial_1-\partial_3)(\partial_1-\partial_4)(\partial_2-\partial_3)(\partial_2-\partial_4)(\partial_3-\partial_4)}{(x_1-x_2)(x_1-x_3)(x_1-x_4)(x_2-x_3)(x_2-x_4)(x_3-x_4)}\airy(x_1)\airy(x_2)\airy(x_3)\airy(x_4).
\end{align}
The partial derivatives generate a total of $64$ terms, of which some cancel, but $24$ remain. For example the first term we would write down is:
\begin{equation}
  \average{\psi(x_1)\psi(x_2)\psi(x_3)\psi(x_4)}\supset \frac{\airy'''(x_1)\airy''(x_2)\airy(x_3)\airy(x_4)}{(x_1-x_2)(x_1-x_3)(x_1-x_4)(x_2-x_3)(x_2-x_4)(x_3-x_4)}.  
\end{equation}
Each such term has $6$ derivatives to distribute among the Airy functions, with a maximum of $3$ per Airy. Taking $x_1\to x_2=x$ and $x_3\to x_4=y$, one ends up with terms as:
\begin{equation}
    \average{\psi^2(x)\psi^2(y)}\supset \frac{1}{(x-y)^4}\airy''''(x)\airy''(x)\airy''(y)\airy(y).
\end{equation}
Each term has now $8$ derivatives to distribute among the Airy functions, with a maximal of $4$ per Airy. Repeatedly applying the Airy equation, one finds after what is very much a bookkeeping exercise:
\begin{equation}
    \average{\psi^2(x)\psi^2(y)}=\frac{1}{(x-y)^2}\average{\psi^2(x)}\average{\psi^2(y)}-\frac{1}{(x-y)^2}K(x,y)^2,
\end{equation}
with $K(x,y)$ the well-known Airy kernel:
\begin{equation}
    K(x,y)=\frac{\airy'(x)\airy(y)-\airy(x)\airy'(y)}{x-y}.
\end{equation}
This replaces the role of the sine kernel $S(E_i,E_j)$ for GUE away from the spectral edge, also in higher cluster functions \cite{mehta}.
\\~\\
Now that we have the appropriate clusters near the spectral edge, we can redo the analysis of section \ref{sect:fixing} and fix eigenvalues in this region, as formulas \eqref{densityfree} and \eqref{twolevelfree} are completely general. For example:
\begin{equation}
    \average{\rho(E)}_\lambda =\delta(E-\lambda) R(E)-\frac{K(E_1,E_2)}{R(\lambda)},\label{exactdepl}
\end{equation}
with $R(E)$ from \eqref{rairy}. For an eigenvalue not too close to the spectral edge, we have:
\begin{equation}
    \average{\rho(E)}_\lambda =\quad\raisebox{-15mm}{\includegraphics[width=50mm]{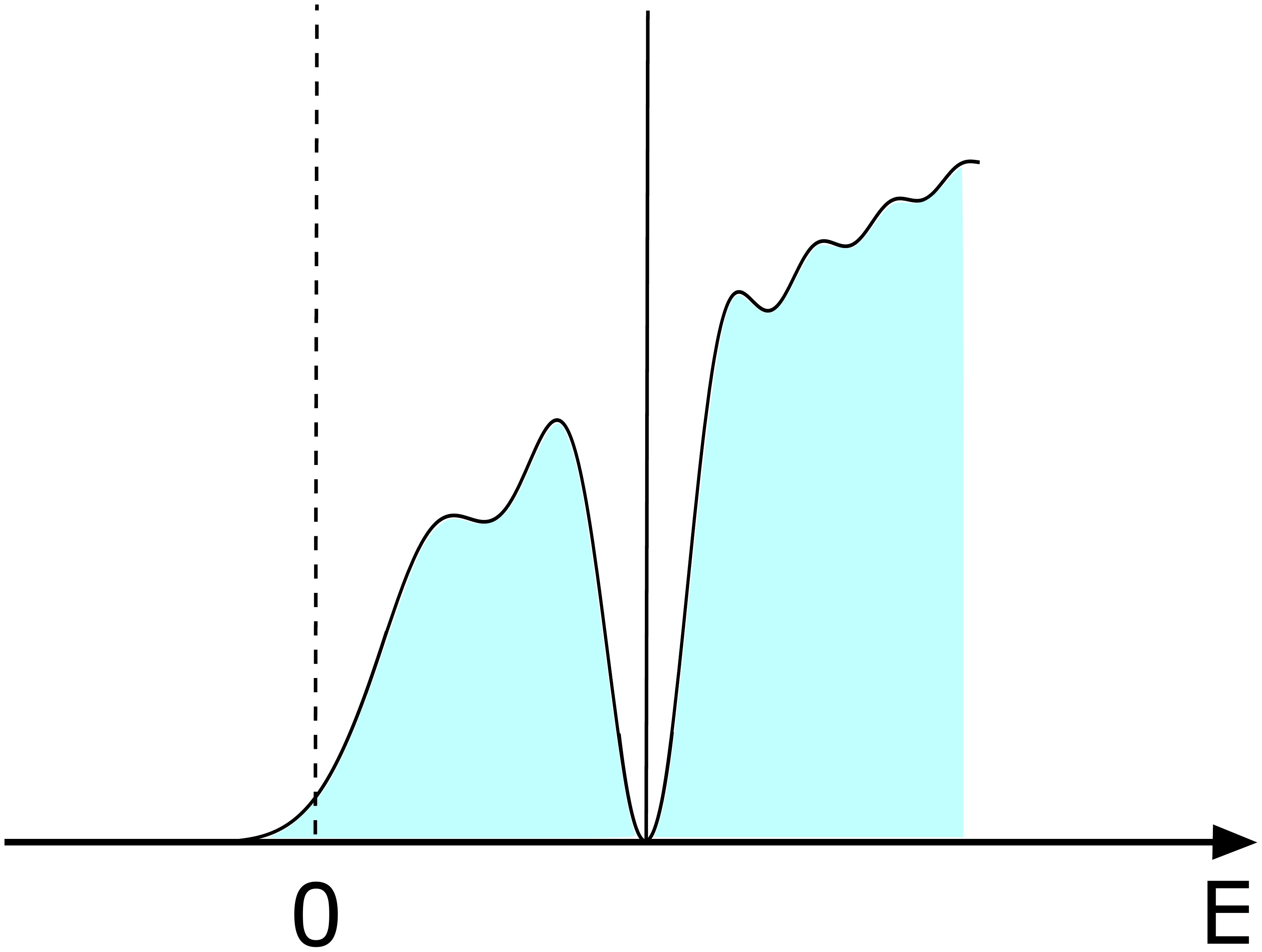}}
\end{equation}
One recognizes the same features as in \eqref{1fixed}. 
\\
It is interesting to see what happens when we insert an eigenvalue very close to the spectral edge or even in the forbidden region $E<0$. Note that this is a very unlikely situation since the total spectral density in the forbidden region is much less than one. Hence, when inserting an eigenvalue in the forbidden region, we expect a depletion in the continuum of essentially the entire forbidden region and of the region closest to the spectral edge. Armed with our exact Airy formula \eqref{exactdepl} we find this is indeed the case. For example when fixing an eigenvalue at the origin, one finds:
\begin{equation}
   \average{\rho(E)}_\lambda =\quad\raisebox{-15mm}{\includegraphics[width=50mm]{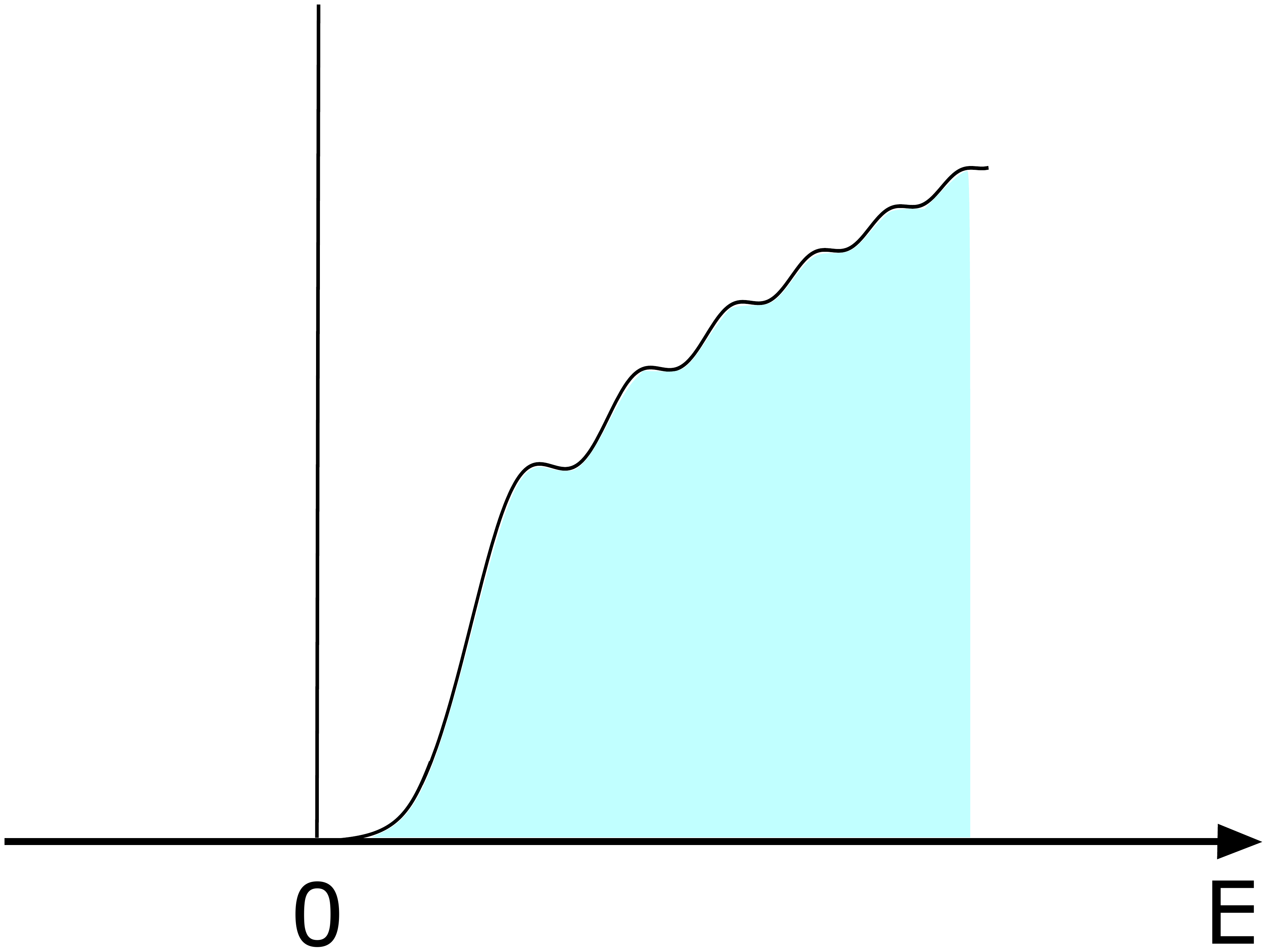}}
\end{equation}
\section{Normalization property}\label{app:add}
One consequence of \eqref{meassuredecompose} is the following:
\begin{align}
    \int_\mathcal{C} d\lambda \, \Delta(\lambda, \lambda_1\dots\lambda_n) &\average{\psi^2(\lambda)\psi^2(\lambda_1)\dots \psi^2(\lambda_n)}_{\LL-n-1}\mathcal{Z}_{L-n-1}\nonumber\\ 
		&=\Delta(\lambda_1\dots \lambda_n)\average{\psi^2(\lambda_1)\dots \psi^2(\lambda_n)}_{\LL-n}\mathcal{Z}_{\LL-n}.\label{property}
\end{align}
Using \eqref{recursive}, we recognize the definition of the correlators \eqref{rbrane} on the left and the $n$-level correlator on the right:
\begin{equation}
\int_\mathcal{C} d\lambda \frac{R(\lambda,\lambda_1\dots\lambda_n)}{R(\lambda_1\dots \lambda_n)}=L-n.\label{prop}
\end{equation}
Since $R(\lambda)=\average{\rho(\lambda)}$, taking $n=0$ in the above we recover the normalization property:
\begin{equation}
    \int_\mathcal{C}d\lambda \average{\rho(\lambda)}=L.\label{norm}
\end{equation}
We can apply \eqref{property} recursively to find:
\begin{equation}
    \int_\mathcal{C} d\lambda_1\dots d\lambda_n\, \Delta(\lambda_1\dots \lambda_n)\, \average{\psi^2(\lambda_1)\dots \psi^2(\lambda_n)}\mathcal{Z}_{L}=\mathcal{Z}_{L+n}.\label{identity}
\end{equation}
This formula appeared in \cite{0408039}. It means that we can add eigenvalues to an ensemble by introducing pairs of branes $\psi^2(\lambda)$ and integrating out $\lambda$. 

\section{Details on late-time contributions}\label{app:details}
We check that certain contributions to the two level spectral density \eqref{twolevelfree} are irrelevant for the behavior in the plateau region $t>2\pi\rho(E)$. Concerning the terms on the second line of \eqref{twolevelfree}, taking the Fourier transform, we are led to:
\begin{equation}
    \average{S_E(t)}_{\lambda_1\dots}\supset 2\sum_{i=1}^n\int_{\text{bin}(E)} d E \,\cos t(E-\lambda_i) \frac{R(E,\lambda_1\dots)}{R(\lambda_1\dots)}.
\end{equation}
The leading factorizing piece decays as a power law:
\begin{align}
    \average{S_E(t)}_{\lambda_1\dots}\supset 2\rho(E)\sum_{i=1}^n \int_{\text{bin}(E)} d E\,\cos t(E-\lambda_i)=\frac{4\rho(E)}{t}\sum_{i=1}^n \sin t\left(\lambda_i+\frac{N}{2\rho(E)}\right).\label{1ont}
\end{align}
Using \eqref{cluster}, we see that the $E$ dependence of the connected piece is due to terms of the type: 
\begin{align}
    &\int_{\text{bin}(E)} d E\,e^{it(E-\lambda_i)}\, \sinc \pi \rho(E)(E-\lambda_j)\,\sinc \pi \rho(E)(E-\lambda_k)\nonumber\\
    \nonumber &=\rho(E) \int_{-\infty}^{+\infty} d\tau \frac{\sin(t-\tau)(\lambda_i+\dots)}{(t-\tau)}\int d E\, e^{i \tau E} \, \sinc \pi \rho(E)(E-\lambda_j)\,\sinc \pi \rho(E)(E-\lambda_k).
\end{align}
The integral over $E$ gets a function $f(\tau)$ which is finite and has compact support. Because of this we can get the late time behavior by Taylor expanding $(t-\tau)^{-1}=1/t+\tau/t^2+ \dots$. The leading term is of the imaginary part of:
\begin{equation}
    \frac{\rho(E)}{2i t}e^{it (\lambda_i+\dots)}\int d\tau\, e^{-i\tau (\lambda_i+\dots)}\, f(\tau).
\end{equation}
This exists and is finite. The result are contributions with the same type of $t$-dependence as \eqref{1ont}. Concerning the final contribution of \eqref{twolevelfree}, we have a leading contribution which gets the averaged ramp \eqref{averagedramp}. This is negligible beyond the plateau time. The other terms have $E_1$ and $E_2$ dependence either of either type:
\begin{align}
    \nonumber &\sim\sinc\pi\rho(E)(E_1-\lambda_i)\,\sinc \pi\rho(E)(E_2-\lambda_j)\,\sinc \pi\rho(E)(E_1-E_2)\\
    &\sim \sinc\pi\rho(E)(E_1-\lambda_i)\,\sinc \pi\rho(E)(E_2-\lambda_j)\,\sinc \pi\rho(E)(E_1-\lambda_k)\,\sinc \pi\rho(E)(E_2-\lambda_l).\nonumber
\end{align}
These result in corrections of the averaged ramp behavior, but does not contribute significantly beyond the plateau time.

\end{document}